\documentclass[12pt]{article}
\usepackage[utf8]{inputenc}
\usepackage{amsbsy}
\usepackage{amstext}
\usepackage{amssymb}
\usepackage{graphicx}
\usepackage{setspace}
\usepackage{url}
\usepackage{siunitx}
\usepackage{multirow}
\usepackage{booktabs}
\usepackage{mathtools}
\usepackage[pdftex,bookmarks=true,colorlinks=true,citecolor=blue,plainpages=false]{hyperref}
\usepackage[round]{natbib}

\usepackage{amsmath}
\usepackage{color}
\usepackage{bm}

\graphicspath{ {figures/} }

\title{A Structural Model of Business Card Exchange Networks\thanks{We are grateful to Naoki Maejima and Michael Schweinberger for their valuable comments and to the engineers at Sansan DSOC for their feedback and for provisioning the necessary infrastructure.} }
\author{Juan Nelson Martínez Dahbura - Sansan, Inc \\ Shota Komatsu - Sansan, Inc\\ Takanori Nishida - Sansan, Inc \\ Angelo Mele - Johns Hopkins University}
\date{Last updated: \today } 

\newtheorem{assumption}{Assumption}

\newtheorem{proposition}{Proposition}

\begin{document}

\maketitle

\abstract{Social and professional networks affect labor market dynamics, knowledge diffusion and new business creation. To understand the determinants of how these networks are formed in the first place, we analyze a unique dataset of business card exchanges among a sample of over 240,000 users of the multi-platform contact management and professional social networking tool for individuals Eight. We develop a structural model of network formation with strategic interactions, and we estimate users' payoffs that depend on the composition of business relationships, as well as indirect business interactions. We allow heterogeneity of users in both observable and unobservable characteristics to affect how relationships form and
are maintained. The model's stationary equilibrium delivers a likelihood that is a mixture of exponential random graph models that we can characterize in closed-form.
We overcome several econometric and computational challenges in estimation, by exploiting
a two-step estimation procedure, variational approximations and minorization-maximization methods. Our algorithm is scalable, highly parallelizable and makes efficient use of computer memory to allow estimation in massive networks. We show that users payoffs display homophily in several dimensions, e.g. location; furthermore, users unobservable characteristics also display homophily.
}

\section{Introduction}

Encounters are a seed of business. Thomas Edison and Henry Ford became friends after a chat at the convention of the Association of Edison Illuminating Companies in New York. Steve Jobs and Steve Wozniak met through a mutual friend, Bill Fernandez. Needless to say, these encounters eventually turned into great businesses, path-breaking innovations, and new products. Social and business networks that ultimately stem from such encounters create interesting economic phenomena, which has attracted many scholars to explore the subject. Indeed, even if they do not cause the birth of new businesses, professional networks play an important role in various economic activities at both individual and firm levels.\footnote{Professional networks provide information about vacancies and quality of job applicants through referrals in labor markets \citep{IoannidesLoury2004,CalvoArmengol2004, Calvo-ArmengolJackson2004, GaleottiMerlino2014, Galenianos2014} as documented in several empirical studies \citep{Beaman2012, BayerRossTopa2008}. Firm-to-firm networks are suggested to catalyze aggregate fluctuations \citep{Acemoglu2012network}, performance heterogeneity among firms \citep{Bernard2019production}, agglomeration \citep{Miyauchi2021}, and knowledge creation and diffusion \citep{Konig2018endogenous}.} However, most of the existing studies presume that there are already more-or-less established relationships among firms or persons, and there are no many empirical studies about how these professional networks are formed in the first place.

In this paper, we employ a unique dataset with over $240,000$ individuals and the $670,000$ business connections among them to estimate a business network formation model, accounting for observable and unobservable individual characteristics that affect the willingness to form professional relationships. The data is a subset of the social network formed by users of Eight, a multi-platform contact management and professional social networking tool for individuals, provided by the Japanese company Sansan, Inc. In this social network, users connect with each other by exchanging business cards, which allows us to analyze a network of mostly \emph{face-to-face} connections across a diverse spectrum of industries, occupations and locations in the whole Japan, on a scale that has not been used in previous work in network economics.

The Japanese labor market is an ideal setting to study the very beginning of business networks. Indeed, business cards are extensively used in Japan as a way of self-introduction, information sharing and establishing business relationships. One nature of business card exchanges is that when two persons exchange business cards, it is very likely that they are meeting each other for the first time. This aspect is different from other social networks such as Facebook and LinkedIn, where people are often acquainted with each other before they become connected on the platform. The fact that business card exchanges in many cases take place at the first meeting allows us to answer the question of how business networks emerge to begin with.

Our approach overcomes many estimation and empirical challenges posed by the scale of the network by carefully using the model's equilibrium implications as well as new and improved algorithms for estimation.
We provide a theoretical framework for understanding face-to-face professional networking, where
agents have observable and unobservable characteristics that affect their willingness to form professional connections. Their payoffs are also affected by link externalities such as popularity or common business connections. The equilibrium of the model provides the likelihood of observing a particular network of professional relationships at a particular point in time, that we use as likelihood of the data \citep{Mele2017, Mele2020, MeleZhu2021, BoucherMourifie2017}. Unobserved heterogeneity is modeled as grouped random effects, thus providing a mixture model of network formation in equilibrium.
We exploit a specification of the model with local externalities, inducing a likelihood that factorizes in between- and within-blocks contributions. This in turn reduces the computational challenge because links across unobserved types/blocks are conditionally independent.

To estimate the structural model with the massive Eight dataset, we develop a scalable two-step estimation algorithm, improving methods from \cite{VuEtAl2013} and \cite{BabkinEtAl2020} and including observable covariates. In the first step, we recover the unobservable heterogeneity by approximating the likelihood of the model with a stochastic blockmodel, thus abstracting from externalities within blocks. As shown in \cite{BabkinEtAl2020}, this approximation works as long as the network is large and the number of unobservable agents' types (the size of the support for the random effect) is relatively large. We derive mean-field variational approximations for the likelihood of the model, and use an expectation-maximization algorithm to estimate the block structure \citep{WainwrightJordan2008, Bishop2006, BickelEtAl2013, BabkinEtAl2020}. Furthermore, we use a minorization-maximization algorithm, which speeds up computation by several magnitudes with respect to standard maximization \citep{BabkinEtAl2020, VuEtAl2013}. Our improved algorithm makes extensive use of the sparsity of the network and efficient sparse matrix algebra routines, as well as a scalable initialization algorithm \citep{Rosvall_2009}, in order to further reduce the memory and time requirements for estimation. In the second step of the algorithm, we estimate the structural payoff parameters using a flexible pseudolikelihood estimator \citep{BoucherMourifie2017}, conditioning on the estimated unobserved heterogeneity in the first step. This two-step procedure allows us to obtain reliable estimates in such a complex model using a large sample.

In the empirical implementation we allow parameters to be a function of between- and within-blocks memberships. We control for homophily in the location, industry and occupation of the users. Our results show that there is homophily in observable characteristics and users prefer to network with users in the same industry-occupation and location, \emph{other things being equal}. We also find that users respond to popularity and tend to form and maintain  links  to popular users as well as users that have common connections. 

Our estimated model can be used to improve the quality of recommendation systems and for counterfactual policy simulations of business networks, e.g. to assess the impact of new services or exogenous events on the networking in equilibrium \citep{Mele2020AEJPol}. \\

We contribute to the network economics literature in three complementary ways.\footnote{For an extensive literature review please refer to \cite{Jackson2008, GrahamDePaula2020, DePaula2017, Chandrasekhar2016} .}
First, we use a tractable structural model of network formation to understand individual and aggregate networking on the job, where agents payoffs depend on observable and unobservable characteristics, as well as linking externalities \citep{Mele2017, Mele2020, MeleZhu2021, BoucherMourifie2017, GrahamDePaula2020}. The network formation process converges to a stationary equilibrium, corresponding to a mixture of exponential random graphs \citep{SchweinbergerHandcock2015, Mele2020, BabkinEtAl2020}.   

Second, we use unique data from a business card exchange platform to estimate the model, in particular preferences for networking that depend on observables, unobservables and endogenous equilibrium network features (equilibrium externalities). Our data contain digitized information about \emph{in-person} business interactions, while most of the literature relies on in-person data collected through surveys,\footnote{A very popular dataset is Add Health, containing a survey of high school friendship networks. Many authors use this dataset, e.g. \cite{Mele2020,BoucherMourifie2017}. Similar datasets are collected in development economics, e.g. \cite{BanerjeeEtAl2013}.} or data from online platforms, where interactions occur exclusively in the online media.

Third, we propose a scalable estimation algorithm for this class of models, by mixing several approximation and estimation methods.  Most of the literature on structural network formation models has relied on small networks to estimate the models, because of the complexities of computing equilibria and the presence of externalities in the model of network formation. In the current work, we also include unobserved heterogeneity in the model, thus increasing the computational complexity even further. Previous work  has approached estimation in different ways. Some authors do not include externalities \cite{Graham2014}, some use equilibrium properties and subnetworks to reduce the computational burden \citep{Sheng2020, DePaulaEtAl2014}, others exploit pseudolikehood methods \citep{BoucherMourifie2017} or incomplete information \citep{Leung2015}.  
We exploit the fact that in equilibrium our model generates networks in the class of hierarchical exponential random graph models \citep{SchweinbergerHandcock2015, Mele2020}, a mixture model of network formation with complex dependencies among links.
Estimation of such models via Bayesian methods is intractable for large networks \citep{SchweinbergerHandcock2015, Mele2020, SchweinbergerStewart2020, BabkinEtAl2020}. Maximum likelihood estimation also does not scale well with the size of the network.
We use variational approximations \citep{WainwrightJordan2008,Bishop2006, BickelEtAl2013} and efficient algorithms to estimate the unobserved heterogeneity; this is achieved by using state-of-the-art computational algorithms \citep{VuEtAl2013, BabkinEtAl2020}, reducing the memory usage and with appropriate initialization of the maximization \citep{Rosvall_2009}.
The second step of the algorithm uses pseudolikelihood methods, conditional on the estimated latent block structure to estimate the structural payoff parameters \citep{BoucherMourifie2017}. Our two-step procedure is similar to ideas proposed in empirical industrial organization or recent work by \cite{BonhommeLamadonManresa2019} for bipartite networks.

The remainder of the paper is organized as follows. In section \ref{section:data} we briefly describe the data from Eight. Section \ref{section:model} develops and analyzes the theoretical model. Section \ref{section:estimation} describes our two-steps estimation algorithm, and results are shown in Section \ref{section:results}. Section \ref{section:conclusion} concludes. Additional details about the computations are provided in appendix \ref{appendix:computation}.

\section{Data}\label{section:data}

We employ data from Eight,\footnote{\url{https://8card.net/en}} a multi-platform contact management and professional social networking tool for individuals launched in 2012 and provided by Sansan, Inc. Founded in 2007, Sansan, Inc. is a Japanese company that offers business card-based services for corporations and individuals. It is the largest provider in the Japanese market, with its corporate service holding over 80\% of the market share.\footnote{\url{https://ir.corp-sansan.com/en/ir/news/news391044820607604029/main/0/link/Presentation\%20Material\%20for\%20FY2020\%20Q2\%20(EN)_revision.pdf}} With more than 2.8 million registered users, Eight is the leading professional social network in Japan. Centered on business cards, it acts as a contact manager, as well as a networking tool, and offers functionality such as a home feed, user profiles, and instant messaging. Users \emph{connect} within the context of the Eight network by scanning each other’s business cards or by sending online friendship requests. Other services by Eight include paid premium plans for individuals and companies, and the direct recruiting platform Eight Career Design.

Eight allows users to scan business cards with a smartphone’s camera, and to set the date the encounter happened\footnote{Some information in English about the business card database can be found in \url{https://datalp.sansan-dsoc.com/}}. OCR algorithms extract information from the business card image, including the name and company of the person and the office address, among other items. This makes it possible to identify individuals and organizations.\footnote{A gentle introduction in English to the digitization process can be found in \url{https://en.sansan-dsoc.com/data/imagerecognition/}} In order to register for the service, users need to scan their own business cards (hereafter called \emph{profile card}). In case of changes, users can update their profile card information by a simple scan.

The data from Eight that is used in this research includes only anonymized information on connections formed between January and December of 2019 among users that have agreed with Eight's Terms of Service. Nodes represent Eight users who have uploaded a profile card at least once by the end of 2019. We keep only nodes for which all covariates used in the analysis have non-missing values and that belong to the largest connected component of the resulting network. An edge is formed between user A and user B when either A scans B’s business card or vice versa. We exclude purely digital connections, and concentrate only on face-to-face encounters. We assume that a business card exchange is bilateral, and therefore the network is undirected. We also consider only the first contact between a pair of users, so that each link has a weight of 1. Self-loops are excluded.

We obtain node attributes from the latest profile card uploaded by the user in order to cover three sources of homophily: geographic proximity, job type similarity and industrial similarity. We extract the user's job category from the job description in its profile card, and assign users an industrial category based on the user's place of employment. We construct an industry-occupation covariate as the interaction between the industrial category of the user's company and the user's occupation code, which can take any of 8,006 unique values.

For measuring geographic proximity we employ the Hexagonal Hierarchical Spatial Index (hereafter H3) created and open sourced by Uber\footnote{\url{https://eng.uber.com/h3/}}. It is an indexing system that projects the sphere of the Earth into an icosahedron and constructs a grid of nested hexagons, each one of which is assigned a unique identifier or index. The H3 indexing system supports 15 resolutions, where hexagons at higher resolutions have a smaller average area\footnote{A table with the mean area per hexagon at each resolution can be found at \url{https://h3geo.org/docs/core-library/restable/}}, and hexagons within the same resolution do not overlap.

In order to assign an H3 index to a node we perform geocoding on the office address in the user's profile card and obtain its latitude and longitude coordinates. The geocoding mechanism is based on data obtained through the Location Reference Information Download Service \footnote{Location Reference Information Download Service (Geospatial Information Authority of Japan) \url{https://nlftp.mlit.go.jp/index.html}}, and address geolocation data \footnote{Address Geolocation Data (Geospatial Information Authority of Japan) \url{https://www.gsi.go.jp/kihonjohochousa/jukyo_jusho.html}}, both provided by the Geospatial Information Authority of Japan. Addresses within Japanese cities are represented using three nested subdivisions: \textit{chome}, \textit{ban} and \textit{gou} (building frontage), in order of granularity. The quality of geocoding varies depending on the availability of data at each region and the level of granularity. Among the addresses in the dataset, 52\% could be matched up to the \textit{gou} level (highest level of accuracy), and 43\% to the \textit{ban} level. The remaining 4.6\% is matched at the \textit{chome} level. We assign each node the H3 index that contains its coordinates at the resolution of 8. The user's location is thus represented by a hexagon of roughly 0.74 $\text{km}^2$. The choice of the resolution represents a trade-off between the capability to model homophily, and memory requirements of its usage for block recovery. Choosing a resolution that is too low or too high would prevent us from capturing the effect of spatial homophily, as too few/many nodes would be located in the same tile, and too low a resolution imposes high memory requirements to the matrix representation of homophily in this dimension.

One alternative would be to perform matching at the zip code level; however, regions sharing the same zip code can differ greatly in area, and important differences may arise between large cities and the countryside. At resolution 8, H3 index similarity captures more business connections than zip code similarity while still being sparse enough for keeping the computation manageable. Although H3 index similarity does not provide a measure of spatial homophily at large distances, the area of tiles at resolution 8 is large enough to contain several high rise office buildings and commercial areas, and therefore captures the cost of business connections at a local level. The H3 index covariate in the dataset has 28,392 unique values.

The resulting network has $242,223$ nodes and $682,920$ edges. The network is very sparse, with a density of roughly \num{2.3e-5}. The network contains $27,289$ triangles and $9,661,321$ 2-stars. All the 47 prefectures and 1,650 cities (roughly 96\% of the total Japanese cities), are represented in the sample. Nodes based in Japan’s largest urban areas in Tokyo and Osaka account for 50\% of the nodes. The data is highly geographically concentrated. The most common H3 index is shared by 2,603 nodes, and 90\% of the tiles contain 10 or less nodes.


34.3\% of the sample is composed by persons in Sales-related occupations, followed by company directors (13.1\%). 61.5\% of the sample holds the ranks of staff. Nodes are more evenly distributed across industrial categories, with the largest category, IT companies, accounting for only 3.3\% of the sample. Same industry-occupation connections represent a 2.3\% of the total connections, while same H3 index connections represent roughly 1.8\%.

\begin{figure}[h]
    \centering
    \caption{Degree distribution}
    \includegraphics[width=70mm]{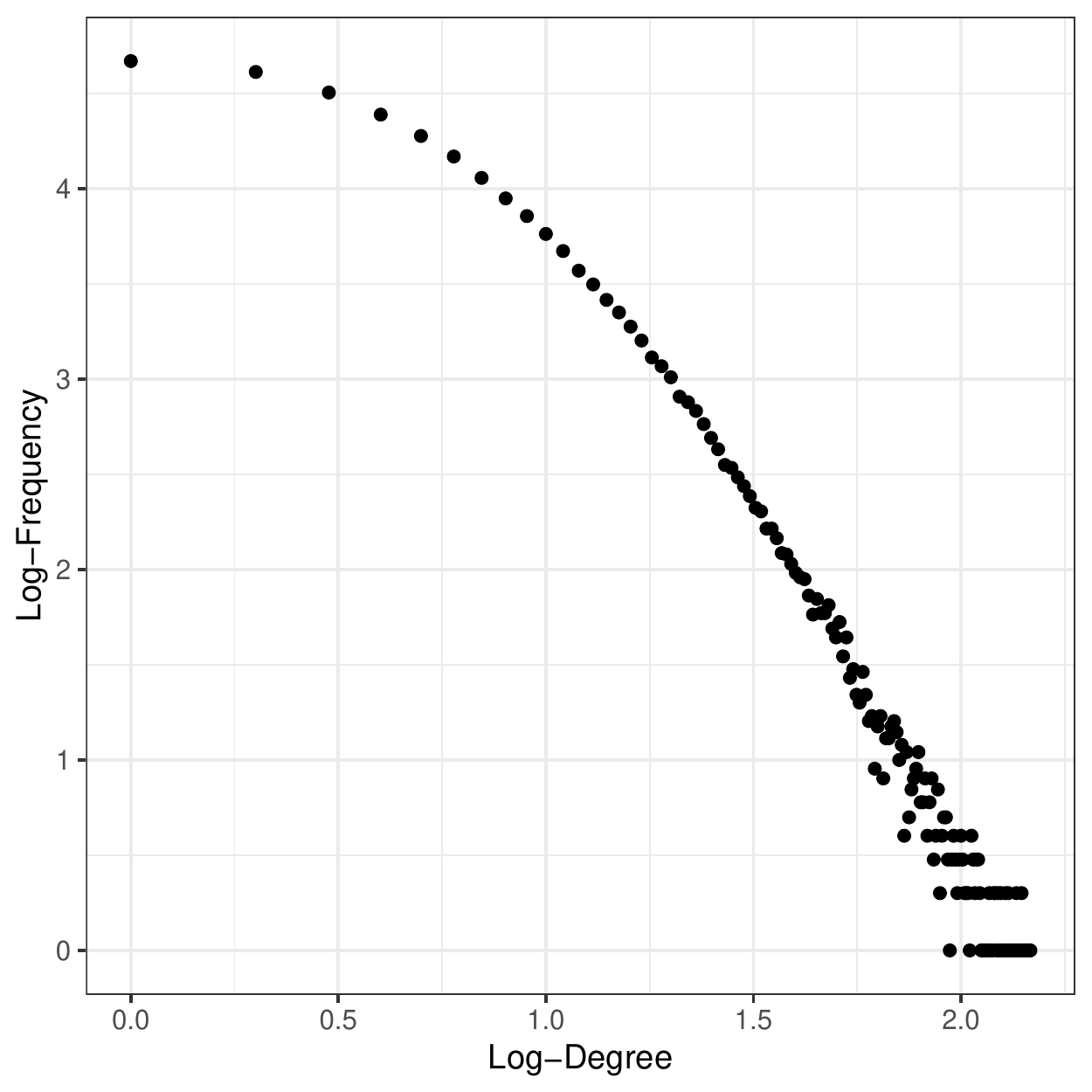}\\
    \begin{quote}
    \scriptsize 
    Note: The figure displays the degree distribution of the network in the data used for estimation. Authors' calculations based on Eight dataset. The network has  $242,223$ nodes and $682,920$ edges, with a density of \num{2.3e-5}. \end{quote}
    \label{fig:degree_dist}
    
\end{figure}

The degree distribution resembles a power law, just like many other large social networks, as shown in Figure \ref{fig:degree_dist}. 

\section{Model}\label{section:model}

We model the decision of users to create professional relationships through in-person interactions or via the business card exchange platform. The set of users is $\mathcal{I}=\lbrace 1, 2, ..., n \rbrace$ and each user $i\in \mathcal{I}$ is characterized by a vector of observable characteristics $\bm{x}_i$, such as gender, location, etc. Additionally, each user is characterized by a $K$-dimensional vector $\bm{z}_i$ that is unobservable to the researcher, but it is observed by other users. The vector $\bm{z}_i = (z_{i1}, ..., z_{iK}) $ is interpreted as an assignment to one of $K$ types; we say that user $i$ belongs to type $k$ if $z_{ik}=1$ and $z_{i\ell} = 0$ for all $\ell\neq k$.

Business card exchanges are recorded in the adjacency  matrix $\bm{g}$, whose generic element $g_{ij}=1$ if users $i$ and $j$ have exchanged a business card, and $g_{ij}=0$ otherwise.

We model users' objective function as a function of the network $\bm{g}$, observable characteristics $\bm{x}$, unobservable types $\bm{z}$ and parameter vector $\bm{\theta} = ( \bm{\alpha}, \bm{\beta},\bm{\psi},\bm{\gamma})$

\begin{eqnarray}
U_i \left( \bm{g},\bm{x},\bm{z};\bm{\theta} \right) = \sum_{j=1}^n g_{ij}u_{ij}(\bm{\alpha},\bm{\beta}) +  \sum_{j=1}^n \sum_{r\neq i,j}^n  g_{ij}g_{jr}w_{ijr}(\bm{\psi}) + \sum_{j=1}^n \sum_{r\neq i,j}^n  g_{ij}g_{jr}g_{ri}v_{ijr}(\bm{\gamma})
\end{eqnarray}
The payoff of direct interactions  $u_{ij}(\bm{\alpha},\bm{\beta}):=u(\bm{x}_i,\bm{x}_j,\bm{z}_i,\bm{z}_j;\bm{\alpha},\bm{\beta})$ includes both costs and benefits of interacting. The payoff is a function of observable characteristics $(\bm{x}_i, \bm{x}_j)$ as well as unobservable types $(\bm{z}_i, \bm{z}_j)$. User $i$ receives a net benefit $u_{ij}(\bm{\alpha},\bm{\beta})$ for a link to user $j$. The second part of the payoff is the effect of popularity $w_{ijr}(\bm{\psi}):=w(\bm{x}_i,\bm{x}_j,\bm{x}_r,\bm{z}_i,\bm{z}_j,\bm{z}_r;\bm{\psi})$. If user $i$ forms a link to $j$, she receives an indirect payoff $w_{ijr}(\bm{\psi})$ from each link formed by $j$. Therefore we can interpret the second term in the utility function as a weighted payoff from popularity, where the weights are functions of observable and unobservable characteristics. Finally, the third term in the payoff is the effect of transitivity $v_{ijr}(\bm{\gamma}):=v(\bm{x}_i,\bm{x}_j,\bm{x}_r,\bm{z}_i,\bm{z}_j,\bm{z}_r;\bm{\gamma})$, or the payoff from common connections. Each user $i$ receives a payoff $v_{ijr}(\bm{\gamma})$ from each user $r$ that is connected to both $i$ and $j$. In the standard strategic network formation literature, direct connections are assumed costly, while indirect connections are free. In this model we do not need to assume that, as the payoff structure allows for costly indirect benefits as well, in principle.

We conceptualize the network formation process as a sequential game, where users form links over time; however, the researcher only observes the network at a particular point in time.\footnote{There is a growing literature in network econometrics using sequential network formation to improve tractability and obtain an equilibrium selection rule. See \cite{DePaula2017, ChristakisEtAl2010, Graham2020, GrahamDePaula2020, Chandrasekhar2016, Mele2017,Mele2020,MeleZhu2021, Jackson2008, JacksonWatts2001} for examples. } We thus focus on the analysis of stationary equilibria of the model. 

In each period two users meet and decide whether to exchange their business cards, by maximizing the surplus generated by the interaction. Before making a decision the users also observe the matching quality of their link, which is also unobserved by the researcher.
Formally, in period $t=0$, each user is randomly assigned to a unobservable type $z_i$, drawn from a Multinomial Distribution
\begin{eqnarray}
      \bm{Z}_i    \overset{iid}{\sim}   Multinomial \left(1;\eta_1, \ldots ,\eta_K\right) 
\end{eqnarray}
Conditional on the realization of the types' assignment $\bm{z}$, the network is formed over time according to the following sequence:

\begin{enumerate}
    \item Two users $i$ and $j$ meet with probability $\rho(\bm{g}_{-ij},\bm{x},\bm{z})$, where $\bm{g}_{-ij}$ is the network $\bm{g}$ excluding the element $g_{ij}$
    \item The users observe a random matching quality shock $\varepsilon_{ij}$
    \item They form a link if the surplus generated by the link is positive, that is if the sum of their payoffs when the link occurs is greater than the sum of their payoffs in absence of a business connection,
    \begin{eqnarray}
          U_i \left( \bm{g}+ij,\bm{x},\bm{z};\bm{\theta} \right) + U_j \left( \bm{g}+ij,\bm{x},\bm{z};\bm{\theta} \right) + \varepsilon_{ij} \geq U_i \left( \bm{g},\bm{x},\bm{z};\bm{\theta} \right) + U_j \left( \bm{g},\bm{x},\bm{z};\bm{\theta} \right)
    \end{eqnarray}
    where the network $\bm{g}+ij$ consists of the network $\bm{g}$ with the addition of the link $g_{ij}$ between users $i$ and $j$. 
\end{enumerate}

This process of network formation generates a Markov Chain of networks, where each network only depends on the previous period's network. In each period, only one link is updated and only two users are actively playing, best-responding to the previous period link decisions of the other players.
To characterize the long-run behavior of the network, we make the following formal assumptions.
\begin{assumption}
 The network formation game satisfies the following assumptions:
    \begin{enumerate}
        \item Users can meet any user with positive probability
        \begin{equation}
            \rho(\bm{g}_{-ij},\bm{x},\bm{z})>0 \text{ \ \ for any } i,j\in \mathcal{I}
        \end{equation}
        and meetings are independent over time and across pairs of users.
        \item The payoffs from popularity and transitivity are invariant to permutation of triads
        \begin{eqnarray}
              w_{ijr}(\bm{\psi}) = w_{\phi(ijr)}(\bm{\psi}) \text{ \ \ and  \ \ } v_{ijr}(\bm{\gamma}) =v_{\phi(ijr)}(\bm{\gamma}) 
        \end{eqnarray}
        for all $i,j,r\in\mathcal{I}$. The notation $\phi(ijr)$ denotes a permutation of the users indicators $ijr$.
        \item The matching quality shock $\varepsilon_{ij}$ is independent and identically distributed over time and across pairs of users according to a logistic distribution. 
    \end{enumerate}
\end{assumption}
The first assumption, imposes that any pair has a positive probability of meeting, however small. This implies that in the long-run two users have (possibly infinitely) many opportunities to form and delete a link. The second assumption restricts the preferences so that we can identify the externality effects. The final assumption about the matching shock is standard in random utility models and is common in the network econometrics literature \citep{GrahamDePaula2020, DePaula2017, Chandrasekhar2016, Graham2020}. 

 Under these assumptions we can show that the game of network formation is a potential game and it converges to a unique stationary equilibrium distribution over networks. Therefore, given the block structure $\bm{z}$, in the long-run we expect to see a network $\bm{g}$ with probability $\pi(\bm{g},\bm{x},\bm{z};\bm{\theta})$, as shown in the next proposition \citep{Mele2017, Mele2020, MeleZhu2021}.
 
 \begin{proposition}\label{prop:potential} 
 Under the assumptions of the model and conditioning on the initial assignment of types $\bm{z}$, the network formation model converges to a stationary Markov Chain of networks, with long-run distribution
 \begin{eqnarray}
       \pi(\bm{g},\bm{x},\bm{z};\bm{\theta}) = \frac{\exp\left[ Q(\bm{g},\bm{x},\bm{z};\bm{\theta}) \right]}{c(\bm{x},\bm{z};\bm{\theta})}
 \end{eqnarray}
 where the potential function $Q(\bm{g},\bm{x},\bm{z};\bm{\theta})$ is 
 \begin{eqnarray}
    Q(\bm{g},\bm{x},\bm{z};\bm{\theta}) &=& \sum_{i=1}^n \sum_{j=1}^n g_{ij}u_{ij}(\bm{\alpha},\bm{\beta}) + \frac{1}{2}\sum_{i=1}^n \sum_{j=1}^n \sum_{r\neq i,j}^n  g_{ij}g_{jr}w_{ijr}(\bm{\psi}) \notag\\
    &+& \frac{2}{3} \sum_{i=1}^n \sum_{j=1}^n \sum_{r\neq i,j}^n  g_{ij}g_{jr}g_{ri}v_{ijr}(\bm{\gamma})   
    \label{eq:potential}
 \end{eqnarray}
 and the normalizing constant $c(\bm{x},\bm{z};\bm{\theta})$ is given by 
 \begin{eqnarray}
       c(\bm{x},\bm{z};\bm{\theta}) =  \sum_{\omega \in \mathcal{G}} \exp\left[ Q(\bm{\omega},\bm{x},\bm{z};\bm{\theta}) \right]
       \label{eq:normalizing_constant}
 \end{eqnarray}
 where $\mathcal{G}$ denotes the set of all possible undirected networks with $n$ nodes.
 \end{proposition}
 The proof can be found in \cite{Mele2020}. 
Proposition \ref{prop:potential} shows that -- after conditioning on the realized unobservable heterogeneity $\bm{z}$ -- the network formation game admits a representation as a potential game \citep{MondererShapley2006}, where all the deterministic incentives of the users to form links are captured by the potential function \eqref{eq:potential}. Indeed, we can show that 
\begin{eqnarray}
& & Q(\bm{g}+ij,\bm{x},\bm{z};\bm{\theta})-Q(\bm{g},\bm{x},\bm{z};\bm{\theta}) \notag \\
& &  =      U_i \left( \bm{g}+ij,\bm{x},\bm{z};\bm{\theta} \right) + U_j \left( \bm{g}+ij,\bm{x},\bm{z};\bm{\theta} \right) - \left( U_i \left( \bm{g},\bm{x},\bm{z};\bm{\theta} \right) + U_j \left( \bm{g},\bm{x},\bm{z};\bm{\theta} \right) \right)
\end{eqnarray}
for all user pairs $i,j\in \mathcal{I}$.\footnote{See \cite{Mele2017, MeleZhu2021, Mele2020} for a formal proof of this statement.} This means that all the incentives of each pair of users to create or delete a link, net of the matching quality, are described by the aggregate potential function. 

The potential game characterization in the proposition implies that the equilibrium pairwise stable networks (with transfers) can be obtained by finding the (local) maxima of the potential $Q(\bm{g},\bm{x},\bm{z};\bm{\theta})$. Therefore, in the long-run we expect to see the pairwise stable networks with high probability, according to the stationary distribution $\pi(\bm{g},\bm{x},\bm{z};\bm{\theta})$.
In such equilibria, the surplus generated by each link is not necessarily split equally between the users involved in the relationship. This allows us to model the fact that some networking relationships are asymmetric or players have different bargaining power \citep{Jackson2008}. 


\section{Estimation}\label{section:estimation}
Estimation of this model is challenging because the likelihood depends on the normalizing constant \eqref{eq:normalizing_constant} that is hard to evaluate even with modern supercomputers \citep{Snijders2002}.\footnote{The constant is the sum of the exponential of potential functions over all possible network configurations. This sum thus includes $2^{(n(n-1)/2}$ terms. Even considering parallelization of the computations, the exact computation of the normalizing constant is either impractical or infeasible for most network sizes. In our data we have around $n=240,000$. Additionally, the standard MCMC methods used in the literature to estimate ERGMs converge too slowly for our data, as in the best case scenario the algorithms converve in $n^2 \log(n)$ steps \citep{BhamidiEtAl2011, Mele2017}. } This is especially true for the size of our dataset. 

To get around some of these challenges, we exploit a particular specification of the model to obtain a likelihood that can be factorized in between- and within-blocks components, after conditioning on the unobserved block structure. This factorization crucially decreases the complexity of computations. 

To estimate the block structure, we approximate the model using a stochastic blockmodel. Given the estimated block structure we estimate the full model using maximum pseudolikelihood estimators. 

These methods bypass the need to compute the likelihood and the normalizing constant, thus allowing estimation in massive networks. In this section we provide several details about the specification of the payoff functions, the estimation of the unobserved heterogeneity (the block structure) and the estimation of the structural payoff parameters.

\subsection{Model specification and likelihood factorization}



The model specification is crucial for a tractable estimation procedure, thus we adopt the following specification with local externalities \citep{Mele2020, Schweinberger2020, SchweinbergerHandcock2015, BabkinEtAl2020}. The parameter $\bm{\alpha}$ only depends on the unobservable types, taking value $\alpha_w$ if the users belong to the same type; otherwise it is $\alpha_b$. The parameters $\bm{\beta}$ contain the net marginal benefits of observable characteristics and we assume that they vary within-types ($\bm{\beta}_w$) and between-types ($\bm{\beta}_b$). Finally, we specify externalities as \emph{local}, that is we assume that the externality is part of the payoff only if all the users involved in the relationship belong to the same unobserved type.
\begin{assumption}\label{assumption:localexternalities} The payoffs of the users are assumed to have the following functional forms:
\begin{eqnarray}
       u_{ij}(\bm{\alpha}, \bm{\beta}) &=& \begin{cases} \alpha_w  + \sum_{p=1}^P \beta_{wp} f_p (\bm{x}_i,\bm{x}_j) & \mbox{if }  \bm{z}_i = \bm{z}_j \\
       \alpha_b  + \sum_{p=1}^P \beta_{bp} f_p (\bm{x}_i,\bm{x}_j) & \mbox{if }  \bm{z}_i \neq \bm{z}_j\end{cases} \\
       w_{ijr}(\bm{\psi}) &=& \begin{cases} \psi & \mbox{if } \bm{z}_i = \bm{z}_j = \bm{z}_r  \\
                                            0 & \mbox{otherwise} \end{cases}\\
       v_{ijr}(\bm{\gamma}) &=& \begin{cases} \gamma & \mbox{if } \bm{z}_i = \bm{z}_j = \bm{z}_r  \\
                                            0 & \mbox{otherwise} \end{cases}\\ 
 \end{eqnarray}
 where the functions $f_p (\bm{x}_i,\bm{x}_j)$ only depend on the observed characteristics of $i$ and $j$, for $p=1,\cdots, P$.
\end{assumption}


The specification differs from other papers using the HERGM framework. In fact, we allow the parameters for the observable covariates to vary within and between blocks, while most papers assume homogeneity for the entire network \citep{SchweinbergerHandcock2015, BabkinEtAl2020}. This allows us more flexibility in estimation. 

The specification with local transitivity and local popularity is convenient for estimation and computation. Indeed, we can show that the potential function (\ref{eq:potential}) can be decomposed in the sum of \emph{within}- and \emph{between}-community potentials. Let $\bm{g}_{k,l}$ denote the sub-network among users in blocks $\mathcal{C}_{k}$ and $\mathcal{C}_{l}$. Let $\bm{x}^{(k)}$ denote the observable covariates of users in community $\mathcal{C}_k$. Let' define the functions:
\begin{eqnarray}
 Q_{k,k}(\bm{g}_{k,l},\bm{x}^{(k)}, \bm{z};\bm{\theta}) &:=& \sum_{i\in \mathcal{C}_k}\sum_{j\in \mathcal{C}_k}g_{ij}u(\bm{x}_i,\bm{x}_j,\bm{z}_i,\bm{z}_j;\alpha_w,\bm{\beta}_w) \\
 &+&  \frac{\psi}{2}\sum_{i\in \mathcal{C}_k}\sum_{j\in \mathcal{C}_k}\sum_{r\in \mathcal{C}_{k}}g_{ij}g_{jr}
 + \frac{2\gamma }{3} 
\sum_{i\in \mathcal{C}_k}\sum_{j\in \mathcal{C}_k}\sum_{r\in \mathcal{C}_{k}}g_{ij}g_{jr}g_{ri} \notag \\
 Q_{k,l}(\bm{g}_{k,l},\bm{x}^{(k)},\bm{x}^{(l)}, \bm{z};\bm{\theta}) &:=& \sum_{i\in \mathcal{C}_k}\sum_{j\in \mathcal{C}_l}g_{ij}u(\bm{x}_i,\bm{x}_j,\bm{z}_i,\bm{z}_j;\alpha_b,\bm{\beta}_b)
\end{eqnarray}

Then the potential function can be re-written as
\begin{eqnarray}
Q(\bm{g},\bm{x},\bm{z};\bm{\theta}) &=& \sum_{k=1}^{K}  Q_{k,k}(\bm{g}_{k,k},\bm{x}^{(k)}, \bm{z};\bm{\theta}) + \sum_{k=1}^{K} \sum_{l>k}^{K} Q_{k,l}(\bm{g}_{k,l},\bm{x}^{(k)},\bm{x}^{(l)}, \bm{z};\bm{\theta})
\label{eq:potential_within_between}
\end{eqnarray}
From a practical standpoint, this decomposition implies that the likelihood of the network factorizes as product of within- and between-community likelihoods, facilitating estimation and identification.
\begin{equation}
\pi(\bm{g},\bm{x},\bm{z};\bm{\theta}) =\prod_{k=1}^{K} \frac{\exp\left[Q_{k,k}(\bm{g}_{k,k},\bm{x}^{(k)}, \bm{z};\bm{\theta})\right]}{c_{k,k}(\mathcal{G}_{k,k},\bm{x}^{(k)};\bm{\theta})}\left[ \prod_{l>k}^{K}\frac{\exp\left[Q_{k,l}(\bm{g}_{k,l},\bm{x}^{(k)},\bm{x}^{(l)}, \bm{z};\bm{\theta})\right]}{c_{k,l}(\mathcal{G}_{k,l},\bm{x}^{(k)},\bm{x}^{(l)};\bm{\theta})} \right]
\label{eq:lik_factorized}
\end{equation}
where the within-community and between-communities normalizing constants are, respectively
\begin{eqnarray}
c_{k,k}(\mathcal{G}_{k,k},\bm{x}^{(k)}, \bm{z};\bm{\theta}) &=& \sum_{\bm{\omega}_{k,k}\in \mathcal{G}_{k,k} }\exp \left[Q_{k,k}(\bm{\omega}_{k,k},\bm{x}^{(k)}, \bm{z};\bm{\theta})\right] \\
c_{k,l}(\mathcal{G}_{k,l},\bm{x}^{(k)}, \bm{x}^{(l)}, \bm{z};\bm{\theta}) &=& \sum_{\bm{\omega}_{k,l}\in \mathcal{G}_{k,l} }\exp \left[Q_{k,k}(\bm{\omega}_{k,l},\bm{x}^{(k)},\bm{x}^{(l)}, \bm{z};\bm{\theta})\right] 
\end{eqnarray}

Notice that the between-community potential $Q_{k,l}(\bm{g}_{k,l},\bm{x}^{(k)},\bm{x}^{(l)}, \bm{z};\bm{\theta})  $ does not include the link externalities (transitivity and popularity). Therefore, the second part of  likelihood (\ref{eq:lik_factorized}) is the product of conditionally independent links,
\begin{equation}
\prod_{l>k}^{K}\frac{\exp\left[Q_{k,l}(\bm{g}_{k,l},\bm{x}^{(k)},\bm{x}^{(l)}, \bm{z};\bm{\theta})\right]}{c_{k,l}(\mathcal{G}_{k,l},\bm{x}^{(k)},\bm{x}^{(l)};\bm{\theta})} = \prod_{l>k}^{K} \prod_{i\in \mathcal{C}_k}\prod_{j\in \mathcal{C}_l}\frac{\exp\left[ g_{ij}\left(u_{ij}(\alpha_b,\bm{\beta}_b) + u_{ji}(\alpha_b,\bm{\beta}_b)\right) \right]}{1+ \exp\left[	 (u_{ij}(\alpha_b,\bm{\beta}_b) + u_{ji}(\alpha_b,\bm{\beta}_b)\right]}
\end{equation}

To summarize, Assumption \ref{assumption:localexternalities} guarantees independence of  between-communities links; on the other hand, within-community links may have strong dependence. In aggregate, our model maintains the complex correlation structure of exponential family random graphs (ERGMs) locally, but allows for weaker dependence among links globally.

\subsection{Estimation algorithm}
The likelihood factorization described in the previous section attenuates some of the computational issues in estimation. However, most applications to date have focused on networks of few hundred nodes when using a Bayesian estimation strategy \citep{SchweinbergerHandcock2015,Mele2020}; and networks with few thousands nodes when using an approximate maximum likelihood strategy \citep{BabkinEtAl2020}. Our data contain hundreds of thousands nodes and therefore we have to use alternative methods and computational strategies to obtain a computationally tractable estimation method.

Our approximate algorithm consists of two steps, as suggested in \cite{BabkinEtAl2020}. In step 1 we estimate the block structure $\hat{\bm{z}}$, approximating the likelihood of the model as the one of a stochastic blockmodel. We then use a variational approximation to obtain a tractable lower bound of the log-likelihood and accelerate the estimation using a minorization-maximization algorithm suggested in \cite{VuEtAl2013}. In step 2, given the estimated block structure $\hat{\bm{z}}$, we estimate the parameters of the model $(\bm{\alpha}, \bm{\beta}, \bm{\psi}, \bm{\gamma})$ using maximum pseudolikelihood estimators.


This procedure is based on two considerations. 
First, the likelihood of a stochastic block-model imposes the same probability of the original likelihood on between-block links. Therefore, the approximation only involves the within-block sub-networks. As long as the network is large, most of the probability mass is on the between-block links, and therefore this approximation works well.\footnote{Formal statements are contained in \cite{BabkinEtAl2020}.} 

Second, while the likelihood of a stochastic block-model is intractable, there exist variational methods of inference to recover its parameters. Variational methods maximize a lower bound to the likelihood, recovering an estimated block structure. The asymptotic analysis shows that variational estimates are consistent and asymptotically normal \cite{BickelEtAl2013, DaudinEtAl2008}.
Computations can be sped up by using  minorization-maximization techniques \citep{VuEtAl2013}.\footnote{Alternatively, spectral methods hold promise in dealing with massive network data \citep{AthreyaEtAl2018a, MeleEtAl2021}. } However, the implementation in \cite{VuEtAl2013} does not take into account the observable covariates, which are crucial in our application. Therefore, we extend their algorithm to include (discrete) covariates. 

We present the two steps in the following subsections, while providing more technical details in Appendix \ref{appendix:computation}.

\subsubsection{Approximate block structure estimation}
In step 1, we approximate the log-likelihood of the model, as if there are no link externalities, i.e. we rewrite the likelihood as if $(\psi, \gamma)= (0,0)$. This approximation works as long as we have many blocks, that is when $K$ is relatively high compared with the size of the network $n$. 

The full likelihood of our model can be written as follows
\begin{equation}
\mathcal{L}(\bm{g},\bm{x};\bm{\theta},\bm{\eta}) = \sum_{\bm{z}\in \mathcal{Z}} L\left( \bm{g}, \bm{x},\bm{z}; \bm{\theta}, \bm{\eta} \right) = \sum_{\bm{z}\in\mathcal{Z}}P_{\bm{\eta}}\left(\bm{Z}=\bm{z}\right) \pi(\bm{g},\bm{x},\bm{z};\bm{\theta}).
\label{eq:complete_likelihood}
\end{equation}
Conditional on the community structure $\bm{z}$, the probability that we observe network $\bm{g}$ is given by $\pi(\bm{g},\bm{x},\bm{z};\bm{\theta})$: this corresponds to the probability of observing the network in the long run, that is
\begin{equation}
\pi(\bm{g},\bm{x},\bm{z};\bm{\theta}) =\prod_{k=1}^{K} \frac{\exp\left[Q_{k,k}(\bm{g}_{k,k},\bm{x}^{(k)}, \bm{z};\bm{\theta})\right]}{c_{k,k}(\mathcal{G}_{k,k},\bm{x}^{(k)};\bm{\theta})}\left[ \prod_{l>k}^{K} \prod_{i\in \mathcal{C}_k}\prod_{j\in \mathcal{C}_l}\frac{\exp\left[ g_{ij}\left(u_{ij}(\alpha_b,\bm{\beta}_b) + u_{ji}(\alpha_b,\bm{\beta}_b)\right) \right]}{1+ \exp\left[	 (u_{ij}(\alpha_b,\bm{\beta}_b) + u_{ji}(\alpha_b,\bm{\beta}_b)\right]} \right]
\label{eq:lik_network}
\end{equation}

The complete likelihood (\ref{eq:complete_likelihood}) is obtained by multiplying the likelihood (\ref{eq:lik_network}) by the probability of firm types/communities $\bm{z}$, that is $P_{\bm{\eta}}\left(\bm{Z}=\bm{z}\right)$, given by a multinomial distribution
\begin{equation}
\bm{Z}_i \vert \eta_1,...,\eta_K  \overset{iid}{\sim} Multinomial\left(1;\eta_1,...\eta_K\right) \text{ for  } i=1,...,n
\end{equation}
and summing over all possible community structures $\bm{z}\in\mathcal{Z}$.\\

Our estimation method is based on the observation that if the externalities are not
present in the model, $\psi=0$ and $\gamma=0$, the likelihood is the same as the one of a 
standard K-block stochastic blockmodel with nodal covariates.
Therefore we consider the approximation
\begin{eqnarray}
L\left( \bm{g}, \bm{x},\bm{z}; \bm{\alpha}, \bm{\beta},\psi,\gamma, \bm{\eta} \right) \approx L\left( \bm{g}, \bm{x},\bm{z}; \bm{\alpha}, \bm{\beta},\psi=0,\gamma=0, \bm{\eta} \right)
\end{eqnarray}
To estimate the block-structure $\bm{z}$ we use a variational approximation for stochastic blockmodels, and 
compute the lower bound of the log-likelihood \citep{WainwrightJordan2008, Bishop2006, BabkinEtAl2020}. 
Let $q(\bm{z})$ be an approximating distribution over blocks $\bm{z}$. Then the lower bound $\ell_B (\bm{g},\bm{x};\bm{\alpha},\bm{\beta}, \bm{\eta})$ is obtained via an application of Jensen's inequality

\begin{eqnarray}
\ell(\bm{g},\bm{x},\bm{\alpha}, \bm{\beta}, \bm{\eta}) &:=&  \log \sum_{\bm{z}\in\mathcal{Z}} L\left( \bm{g}, \bm{x},\bm{z}; \bm{\alpha}, \bm{\beta},\psi=0,\gamma=0, \bm{\eta} \right) \\
&=&\log \sum_{\bm{z}\in\mathcal{Z}} q(\bm{z}) 
\frac{L\left( \bm{g}, \bm{x},\bm{z}; \bm{\alpha}, \bm{\beta},\psi=0,\gamma=0, \bm{\eta} \right) }{q(\bm{z})} \\
&\geq &  \sum_{\bm{z}\in\mathcal{Z}} q(\bm{z}) \log \left[
\frac{L\left( \bm{g}, \bm{x},\bm{z}; \bm{\alpha}, \bm{\beta},\psi=0,\gamma=0, \bm{\eta} \right) }{q(\bm{z})} \right] \\
&\equiv & \ell_B (\bm{g},\bm{x},\bm{\alpha}, \bm{\beta}, \bm{\eta}).
\end{eqnarray}

The variational method finds the best approximating distribution $q(\bm{z})$, by finding the best lower bound. 
Because this variational problem is also intractable and cannot be solved in closed-form, we choose $q(\bm{z})$ 
from a tractable family of distributions, as suggested in the literature \citep{WainwrightJordan2008}. 
For our model, it is natural to choose a multinomial distribution $q_{\bm{\xi}}(\bm{z})$
\begin{equation}
\bm{Z}_i   \overset{ind}{\sim} Multinomial\left(1;\xi_{i1},...\xi_{iK}\right) \text{ for  } i=1,...,n
\end{equation}
that can be optimized with respect to the \emph{variational parameters} $\bm{\xi}_i$'s to
obtain a \emph{tractable} bound $\ell_B (\bm{g},\bm{x},\bm{\alpha}, \bm{\beta}, \bm{\eta}; \bm{\xi})$
\begin{eqnarray}
\ell_B (\bm{g},\bm{x},\bm{\alpha}, \bm{\beta}, \bm{\eta}; \bm{\xi}) & \equiv &
\sum_{\bm{z}\in\mathcal{Z}} q_{\bm{\xi}}(\bm{z}) \log \left[
\frac{L\left( \bm{g}, \bm{x},\bm{z}; \bm{\alpha}, \bm{\beta},\psi=0,\gamma=0, \bm{\eta} \right) }{q_{\bm{\xi}}(\bm{z})} \right] \\
& = & \sum_{i<j}^{n}\sum_{k=1}^{K}\sum_{l=1}^{K} \xi_{ik} \xi_{jl} \log \pi_{ij,kl}(g_{ij}, \bm{x},\bm{z}) + 
\sum_{i=1}^{n}\sum_{k=1}^{K}\xi_{ik}\left( \log \eta_{k}-\log \xi_{ik}\right) \notag
\label{eq:tractiable_lb}
\end{eqnarray}
where $\log \pi_{ij,kl}(g_{ij}, \bm{x},\bm{z})$ is the log-likelihood of a link between nodes in blocks $k$ and $l$
\begin{eqnarray}
\log \pi_{ij,kl}(g_{ij}, \bm{x},\bm{z}) &\equiv &  g_{ij}\log  \left[ \frac{\exp\left[u_{ij,kl}(\bm{\alpha},\bm{\beta}) + u_{ji,lk}(\bm{\alpha},\bm{\beta})\right]}{1+\exp\left[u_{ij,kl}(\bm{\alpha},\bm{\beta}) + u_{ji,lk}(\bm{\alpha},\bm{\beta})\right] } \right] \notag \\
&+ & (1-g_{ij})\log \left[ \frac{1}{1+\exp\left[u_{ij,kl}(\bm{\alpha},\bm{\beta}) + u_{ji,lk}(\bm{\alpha},\bm{\beta})\right] } \right] \notag
\label{eq:sbm_logproblink}
\end{eqnarray}
and $u_{ij,kl}(\bm{\alpha},\bm{\beta})=u(\bm{x}_i,\bm{x}_j,z_{ik}=z_{jl}=1,\bm{z};\bm{\alpha},\bm{\beta})$ is the direct net benefit payoff of user $i$ in block $k$ from forming a link with user $j$ in block $l$. \\

\noindent \textbf{Minorization-Maximization}. This estimation framework for stochastic blockmodels is relatively standard in the literature and it enjoys several asymptotic properties and guarantees \citep{BickelEtAl2013}. In particular, the estimates are consistent and asymptotically normal. The EM algorithm consists of iteratively updating the parameters via an \emph{expectation} and a \emph{maximization} step, whose updates are available in closed-form \citep{DaudinEtAl2008, BickelEtAl2013}.

 However, maximizing the lower bound
$\ell_B (\bm{g},\bm{x},\bm{\alpha}, \bm{\beta}, \bm{\eta}; \bm{\xi})$ with respect to $\bm{\xi}$ can still be impractical in very large datasets, because the iterative update for each $\xi_{ik}$ depends on $(n-1)K$ other terms $\xi_{jl}$. These updates are time consuming. Additionally, the iterative algorithm used for computing the lower bound approximation is a local algorithm and may get stuck in a local maximum.

To alleviate these computational problems, we extend the Minorization-Maximization methods of \cite{VuEtAl2013} in two complementary directions. First we allow the algorithm to incorporate (discrete) covariates. The original algorithm is designed for stochastic blockmodels without any observable covariates, so this is a significant improvement in terms of applicability of the method. Second, we provide an efficient computational algorithm that exploits matrix algebra rather than nested loops in computation, to speed up computations by a factor of 14000. This allows estimation in massive networks. Our implementation takes advantage of the sparsity of the graph by making use of sparse matrices where possible in order to make efficient use of the memory, which is a problem when dealing with massive datasets.

The idea of minorization algorithms is to find  a function that approximates the lower bound $\ell_B (\bm{g},\bm{x},\bm{\alpha}, \bm{\beta}, \bm{\eta}; \bm{\xi})$, while being simpler to maximize. 
In practice, a function $M\left(\bm{\xi}; \bm{g},\bm{x},\bm{\alpha},\bm{\beta},\bm{\eta}, \bm{\xi}^{(s)} \right)$ minorizes the likelihood lower bound $\ell_B (\bm{g},\bm{x},\bm{\alpha}, \bm{\beta}, \bm{\eta}; \bm{\xi})$ at parameter $\bm{\xi}^{(s)}$ and iteration $s$ if 
\begin{eqnarray}
M\left(\bm{\xi}; \bm{g},\bm{x},\bm{\alpha},\bm{\beta},\bm{\eta}, \bm{\xi}^{(s)} \right) & \leq & \ell_B (\bm{g},\bm{x},\bm{\alpha}, \bm{\beta}, \bm{\eta}; \bm{\xi}) \text{ \ \ for all } \bm{\xi} \\
M\left(\bm{\xi}^{(s)}; \bm{g},\bm{x},\bm{\alpha},\bm{\beta},\bm{\eta}, \bm{\xi}^{(s)} \right) & = &\ell_B (\bm{g},\bm{x},\bm{\alpha}, \bm{\beta}, \bm{\eta}; \bm{\xi}^{(s)})
\end{eqnarray} 
where $\bm{\alpha}, \bm{\beta}, \bm{\eta} $ and $\bm{\xi}^{(s)}$ are fixed.

We follow \cite{VuEtAl2013} and use the following function for a stochastic block model lower bound 
\begin{eqnarray}
M\left(\bm{\xi}; \bm{g},\bm{x},\bm{\alpha},\bm{\beta},\bm{\eta}, \bm{\xi}^{(s)} \right) &:= & \sum_{i<j}^{n} \sum_{k=1}^{K} \sum_{l=1}^{K} \left(\xi_{ik}^{2} \frac{\xi_{jl}^{(s)}}{2\xi_{ik}^{(s)}} + \xi_{jl}^{2} \frac{\xi_{ik}^{(s)}}{2\xi_{jl}^{(s)}}\right) \log \pi_{ij;kl}^{(s)} (g_{ij}, \bm{x},\bm{z}) \notag \\
&+& \sum_{i=1}^{n} \sum_{k=1}^{K} \xi_{ik} \left(\log\eta_{k}^{(s)} - \log\xi_{ik}^{(s)} - \frac{\xi_{ik}}{\xi_{ik}^{(s)}} + 1\right).
\label{eq:minorizer}
\end{eqnarray}
The main difference from \cite{VuEtAl2013} is that our model includes observable (discrete) covariates. Therefore, the updates of the maximization are slightly different.

As in a standard Variational EM algorithm, we can write down the parameter updates in closed-form. The update rules for $\bm{\xi}$, $\bm{\eta}$, and $\pi_{ij;kl}(g_{ij}, \bm{x},\bm{z})$ follow
\begin{align*}
\bm{\xi}^{(s+1)} &\coloneqq \arg \max_{\bm{\xi}} 
M\left(\bm{\xi}; \bm{g},\bm{x},\bm{\alpha}^{(s)},\bm{\beta}^{(s)},\bm{\eta}^{(s)}, \bm{\xi}^{(s)} \right),
\end{align*}
\begin{align*}
\eta_{k}^{(s+1)} \coloneqq \frac{1}{n} \sum_{i=1}^{n} \xi_{ik}^{(s+1)}, \quad k = 1, \ldots, K,
\end{align*}
and 
\begin{align*}
\pi_{ij;kl}^{(s+1)}(d, \chi_{1}, \ldots, \chi_{p},\bm{z}) \coloneqq \frac{\sum_{i=1}^{n}\sum_{j \neq i}\xi_{ik}^{(s+1)} \xi_{jl}^{(s+1)} \bm{1}\lbrace g_{ij}=d, \chi_{1, ij}=\chi_{1}, \ldots, \chi_{p, ij}=\chi_{p} \rbrace }{\sum_{i=1}^{n}\sum_{j \neq i}\xi_{ik}^{(s+1)} \xi_{jl}^{(s+1)} \bm{1}\lbrace \chi_{1, ij}=\chi_{1}, \ldots, \chi_{p, ij}=\chi_{p}\rbrace },
\end{align*}
for $ k, l = 1, \ldots, K$ and $d, \chi_{1}, \ldots, \chi_{p} \in \{0, 1\} $,
respectively. In the formula for $\pi_{ij;kl}^{(s+1)}(d, \chi_{1}, \ldots, \chi_{p},\bm{z}))$ we have used the notation $\chi_{p,ij}$ to denote and indicator variable equal to 1 if the (discrete) nodal covariate $p$ of $i$ and $j$ are the same, i.e. $\chi_{p,ij} = \bm{1}\lbrace x_{ip}=x_{jp} \rbrace  $. Generalizations of this specification are allowed.

The estimated block structure $\widehat{\bm{z}}$ is obtained by choosing the modal block assignment, that is
$\widehat{z}_{ik}=1$ if $\widehat{\xi}_{ik} \geq \widehat{\xi}_{i\ell} $ for all $\ell \neq k$ and $\widehat{z}_{il}=0$ for all $l\neq k$.

\subsubsection{Estimation of structural parameters}
Conditioning on the estimate of $\widehat{\bm{z}}$, we estimate the structural parameters $\bm{\theta} = \left(\bm{\alpha}, \bm{\beta},\psi,\gamma \right)$ by maximum pseudolikelihood (MPLE) methods \citep{BoucherMourifie2017, Snijders2002, BabkinEtAl2020}. This amounts to maximize the product of the conditional link probabilities. 

Formally, given the estimated $\widehat{\bm{z}}$, we compute the conditional probability of a link
\begin{eqnarray}
p_{ij}(\bm{g},\bm{x},\bm{\theta};\widehat{\bm{z}}) = \Lambda \left( u_{ij}(\bm{\alpha},\bm{\beta})+u_{ji}(\bm{\alpha},\bm{\beta}) + \psi\sum_{r\neq i,j}(g_{jr}+g_{ir}) + 4 \gamma \sum_{r\neq i,j} g_{jr}g_{ir} \right)
\end{eqnarray}
where $\Lambda(u) = e^u /(1+e^u)$ is the logistic function. The pseudolikelihood function is
\begin{eqnarray}
\ell_{PL}(\bm{g},\bm{x},\bm{\theta}; \widehat{\bm{z}}) &=& \sum_{i=1}^n \sum_{j> i}^n g_{ij}\log p_{ij}(\bm{g},\bm{x},\bm{\theta}) + (1-g_{ij}) \log (1-p_{ij}(\bm{g},\bm{x},\bm{\theta})) 
\label{eq:logPL}
\end{eqnarray}
and the estimator is the maximizer of the log-pseudolikelihood
\begin{eqnarray}
\widehat{\bm{\theta}}_{PL} = \arg\max_{\bm{\theta}} \ell_{PL}(\bm{g},\bm{x},\bm{\theta}; \widehat{\bm{z}})
\end{eqnarray}
The asymptotic framework for the maximum pseudolikelihood estimator was recently analyzed in \cite{BoucherMourifie2017}. It can be shown that the estimates are consistent and asymptotically normal under some regularity conditions.


\section{Results}\label{section:results}
We estimated our model for the network presented in Section \ref{section:data} with a maximum of 1,500 blocks, using 250 iterations of the EM algorithm. After recovering the estimated block structure $\widehat{\bm{z}}$ and controlling for the node covariates described in Section \ref{section:data}, we estimate the structural parameters using the maximum pseudolikelihood method and accounting for node covariates on the block recovery step. For comparison, we also perform the estimation without taking node covariates into account for block recovery. Our implementation of the algorithm used to obtain these results is available in the \texttt{lighthergm} R package, which can be found at \url{https://github.com/sansan-inc/lighthergm}. We present the results of each step in detail in the subsections below.

\subsection{Block structure estimation results}
The block structure estimation step is by far the most computationally intensive part of the estimation. For this application we employed an Ubuntu Linux machine with 128 GB of memory and 64 processor cores. We set the maximum number of types/blocks to 1,500. The computation is performed with about 35 GB of memory for the block recovery step accounting for node covariates, although it can be performed with well under 32 GB of memory when node covariates are not employed. All processor cores are in use during most of the calculation time.

First, we initialize the blocks by using the Infomap algorithm by \cite{Rosvall_2009}. Infomap presents several advantages over other clustering algorithms for our particular use case. First, Infomap's time complexity is linear in the number of edges, which makes it a good choice for initializing the block memberships on very sparse networks. \cite{Yang2016} show that Infomap performs better than other algorithms with similar time complexities at the same values of the mixing parameter. Furthermore, Infomap's performance at recovering the true communities is independent of the network size. In comparison, the default initialization algorithm on the original \texttt{hergm} R package version 4.1-7 is \emph{Walktrap} \citep{Pascal2005}, which, despite having properties that make it a good candidate, has a space complexity that is quadratic in the number of nodes, making it an expensive choice for clustering large networks.\\

\begin{figure}[h]
    \centering
    \caption{Block size distribution}   \includegraphics[width=70mm]{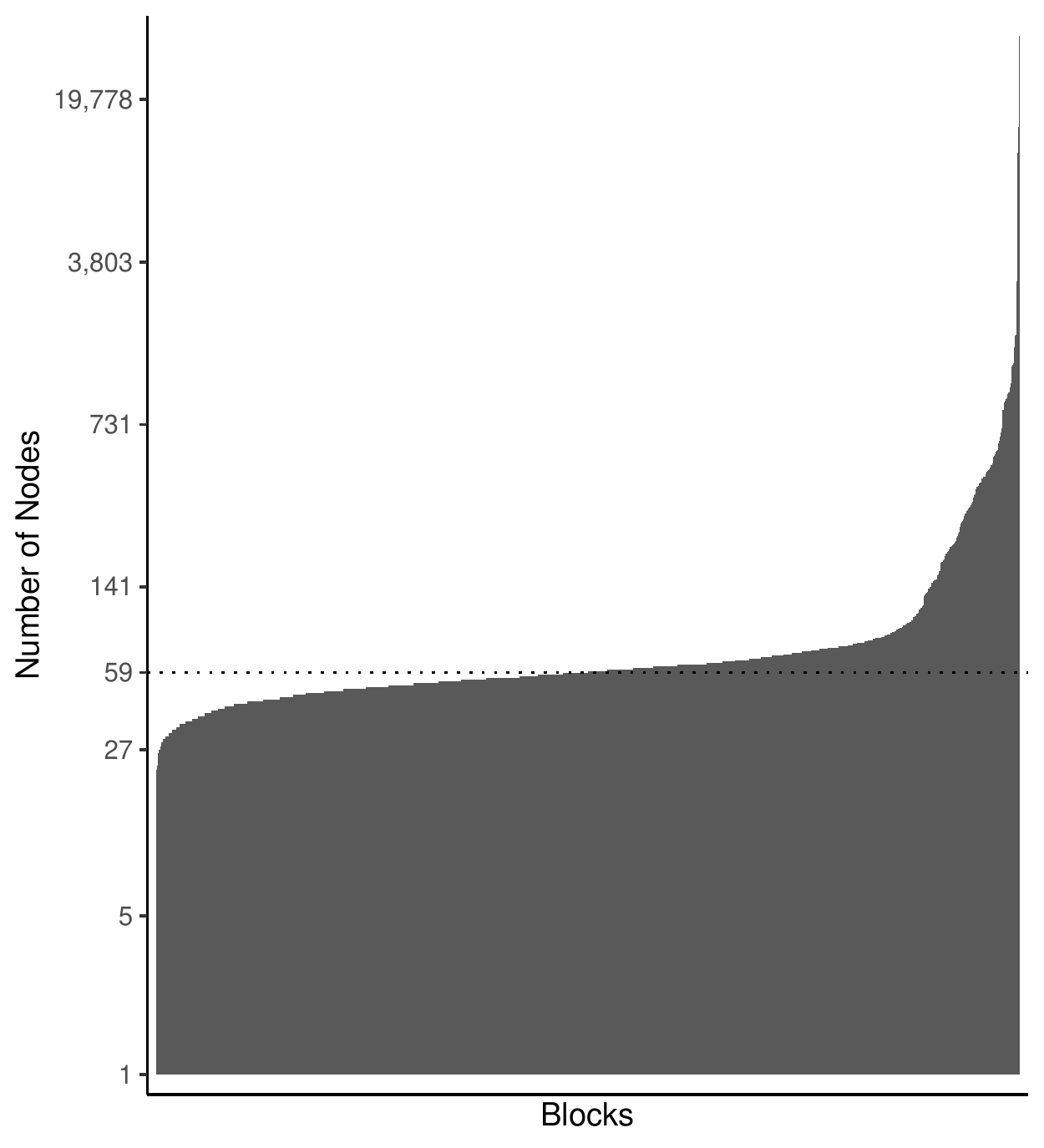}\\
    \begin{quote}
    \scriptsize 
    \label{fig:cluster_size_dist}
    Note: Blocks and their respective sizes are shown in ascending order with respect to the number of affiliated nodes. The vertical axis is expressed in logarithmic scale.\end{quote}
\end{figure}

Starting at the initial block structure, we apply 250 iterations of the fast EM algorithm. Each EM algorithm iteration takes approximately 14 minutes, for a total of 38.3 hours to complete the whole EM iteration part of the block structure estimation step when accounting for node covariates. In comparison, \cite{VuEtAl2013} employ a similar variational approach on a network of $131,000$ nodes with only $20$ blocks and without nodal characteristics with 100 EM iterations taking a total of 24 hours.

\begin{figure}[h]
    \centering
    \caption{Lower Bound Improvement per Iteration}
    \includegraphics[scale=0.7]{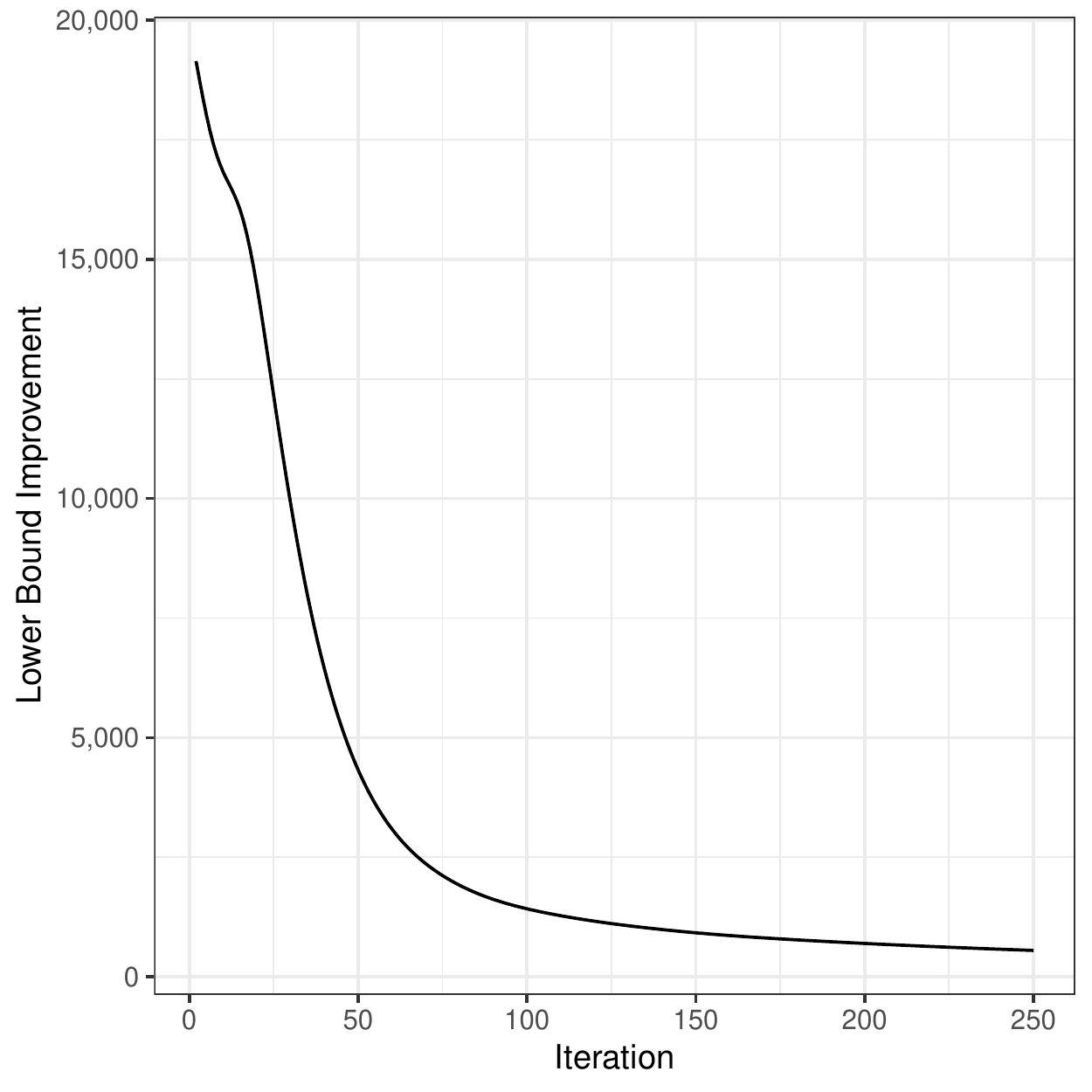}
    \label{fig:lb_improvements}
    \begin{quote}\scriptsize 
    \textit{Note:} Block recovery takes into consideration the information on H3 tile similarity and industry-occupation similarity. \end{quote}
\end{figure}

Figure \ref{fig:lb_improvements} shows the improvement in the target function's Lower Bound at each iteration. The improvements are monotonic, and converge after close to 100 iterations, although the improvement keeps being positive after 250 iterations. In order to understand how much the block structure changes with the number of iterations, we compute the Yule's coefficient with respect to the initial block structure obtained by Infomap. The Yule's coefficient measures the similarity between two block structures regardless of the labels. It takes values between 0 and 1, where higher values mean a higher similarity. We find that the final blocks differ considerably from the initial structure. After 100 iterations, the Yule's coefficient is roughly 0.74, and after 250 iterations it goes down to 0.05.

\begin{figure}[h]
    \centering
    \caption{The business network and the recovered community structure}
    \includegraphics[scale=0.3]{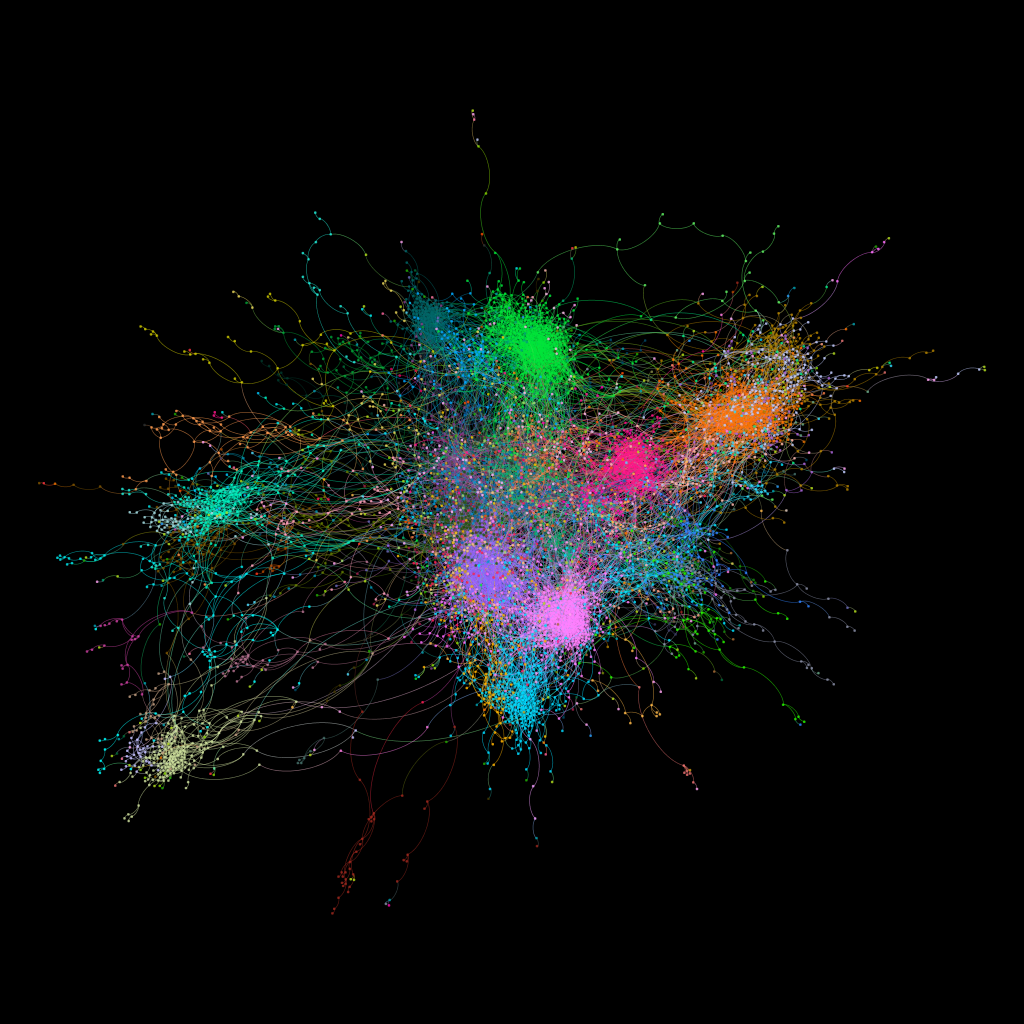}
    \begin{quote}\scriptsize 
    Note: The figure visualizes the network, with colors representing the block membership of the nodes. The visualized network consists of the subnetwork that results from sampling 10,000 nodes from the largest 100 blocks, weighting by the node's degree, and extracting the giant component. The visualization was created with Gephi (\url{https://gephi.org/}). \end{quote}
    \label{fig:network_visualization}
\end{figure}

Figure \ref{fig:cluster_size_dist} shows the sizes of all the blocks in ascending order of the number of affiliated nodes. The dotted line marks the median block size of 59 nodes. We observe that a few blocks contain a large number of nodes, and the largest block contains 37,615 nodes, representing a 15.5\% of the total nodes. Looking at the distribution as a whole, blocks are in general quite homogeneous in size. The block size distribution is strongly concentrated around the median, and has an interquartile range of only 22 nodes. Figure \ref{fig:network_visualization} displays a subset of the network and the estimated block affiliation of the nodes, and shows that the recovered block structure in fact represents areas of the graph with denser connectivity.

\begin{figure}[h]
    \centering
    \caption{Share of the five largest blocks by prefecture}
    \includegraphics[width=100mm]{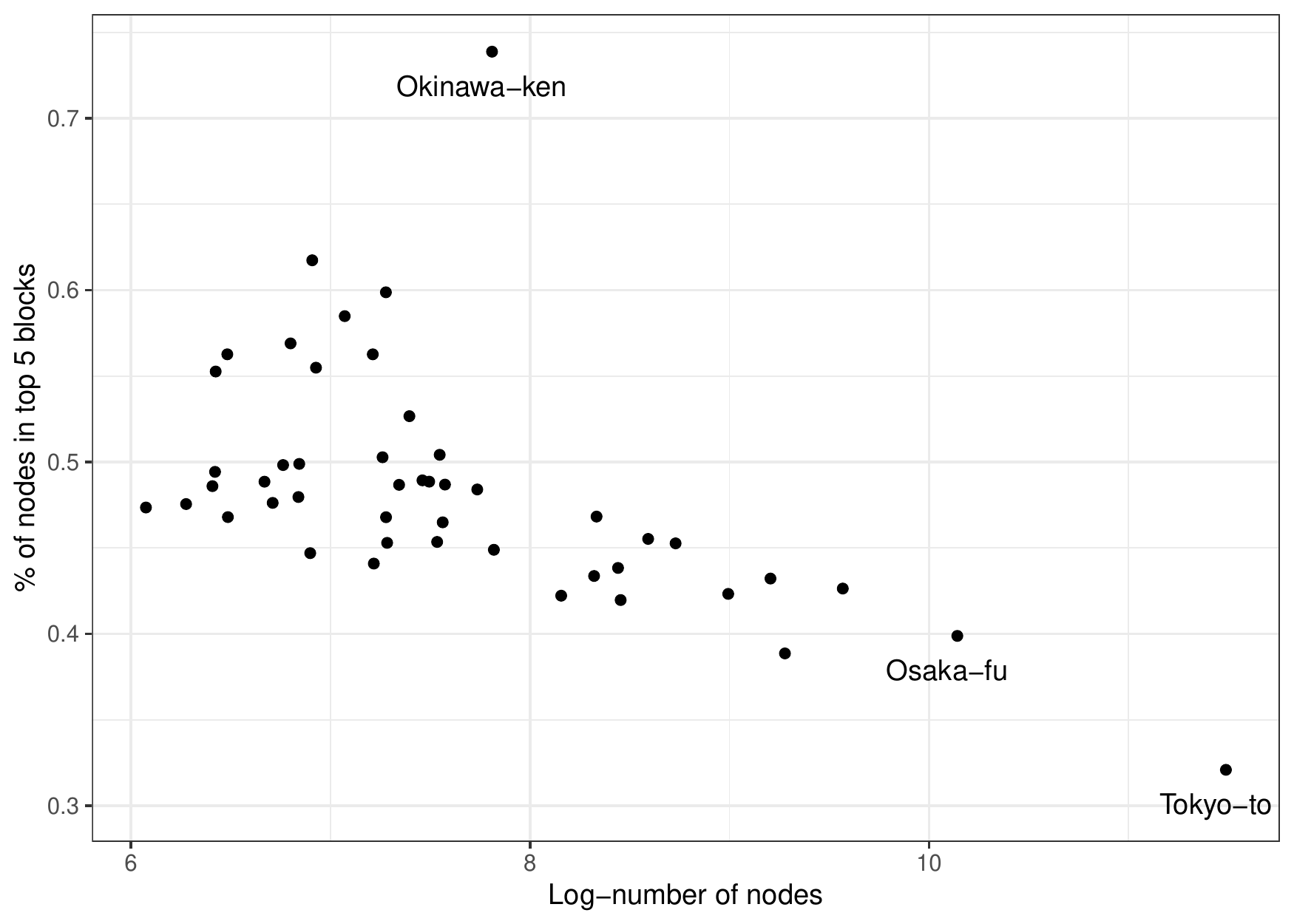}
    \begin{quote}
    \scriptsize
    Note: We measure the diversity of the community structure for each prefecture by ordering blocks in decreasing order by size, and measuring the share of nodes affiliated to the top five blocks. Prefectures with more nodes tend to be more diverse in terms of community affiliation.
    \end{quote}
    \label{fig:top_clusters_by_prefecture}
\end{figure}

Figure \ref{fig:top_clusters_by_prefecture} shows the relationship between the number of nodes and the share of the five largest blocks by prefecture. Nodes in hub prefectures such as Tokyo and Osaka tend to be distributed across many blocks, with no single dominating cluster. On the other hand, smaller prefectures tend to be dominated by a few large blocks. Okinawa is a clear outlier, with the largest block accounting for over 40\% of its nodes.

Finally, when covariates are not employed at the block recovery step, a more skewed distribution of block sizes is obtained. The size of the largest block in this case is 85,512, and the median block size is 72. The Yule's coefficient between this partition and the one obtained when employing node covariates is 0.37, suggesting that in fact, accounting for homophily on observable characteristics can have an impact on the structure of the resulting partition, holding the number of EM iterations and the initial block structure constant.

\subsection{Structural parameters estimation results}
We take the community structure obtained in the previous step and proceed to estimate the within-block and between-block ERGM parameters. Given the assumptions in our model, we separate the between-block connections from the within-block ones, and estimate the parameters for each set independently employing maximum pseudolikelihood estimation. This step requires fewer resources and processing time compared to the previous step. The estimates for both sets of parameters are shown in Table \ref{table:parameter_estimates}.

\begin{table}[!htbp] \centering 
  \caption{HERGM Parameter Estimates} 
  \label{table:parameter_estimates} 
\begin{tabular}{@{\extracolsep{5pt}}lcccc} 
\\[-1.8ex]\hline 
\hline \\[-1.8ex] 
 & \multicolumn{2}{c}{Between} & \multicolumn{2}{c}{Within} \\ 
\cline{2-5} 
\\[-1.8ex] & No Covars & With Covars & No Covars & With Covars\\ 
\\[-1.8ex] & (1) & (2) & (3) & (4)\\ 
\hline \\[-1.8ex] 
 Edges & $-$11.074$^{***}$ & $-$11.000$^{***}$ & $-$9.989$^{***}$ & $-$9.047$^{***}$ \\ 
  & (0.005) & (0.005) & (0.002) & (0.003) \\ 
  & & & & \\ 
 Triangles &  &  & 0.891$^{***}$ & 0.720$^{***}$ \\ 
  &  &  & (0.006) & (0.006) \\ 
  & & & & \\ 
 2-Stars &  &  & 0.111$^{***}$ & 0.100$^{***}$ \\ 
  &  &  & (0.0002) & (0.0002) \\ 
  & & & & \\ 
Same Location (H3 Tile) & 2.130$^{***}$ & 2.049$^{***}$ & 0.989$^{***}$ & 0.793$^{***}$ \\ 
   & (0.041) & (0.041) & (0.027) & (0.028) \\ 
  & & & & \\ 
 Same Industry-occupation & 2.611$^{***}$ & 2.645$^{***}$ & 1.745$^{***}$ & 0.739$^{***}$ \\ 
  & (0.039) & (0.037) & (0.019) & (0.020) \\ 
  & & & & \\ 
\hline \\[-1.8ex] 
Bayesian Inf. Crit. & 37,691,931,592 & 37,691,988,973 & 4,106,060 & 3,129,658 \\ 
\hline 
\hline \\[-1.8ex]
\end{tabular} 
\flushleft \textit{Notes:} $^{*}$p$<$0.1; $^{**}$p$<$0.05; $^{***}$p$<$0.01.\\
\noindent All estimates are obtained using a maximum pseudolikelihood estimator, conditioning on the estimated block structure. Block recovery was performed using 250 EM iterations and 1,500 blocks. For columns (1) and (3) the block structure recovery in the first step does not take into account the covariates; for columns (2) and (4) the first step controls for the covariates.
\end{table}

Standard errors correspond to the ones obtained by each separate maximum pseudolikelihood estimation and thus do not consider the error in the block structure recovery step. Regarding the between-block model results, estimates are similar regardless of whether node covariates were employed to inform the block recovery step. This is to be expected, given that most of the possible connections are across blocks, despite the differences in the final block structure. Coefficients are all significant at the 1\% level. The coefficient for the edges term reflects the sparsity of the network, while geospatial and industrial-occupational homophily are significant factors explaining business connections across blocks.


Results for the within-block connection model show that random connections within the same community are slightly more likely than across blocks, suggesting that persons prefer forming business connections among peers within the same communities, everything else constant, although a formal statistical test is required. We observe a significant preference for transitivity and popularity, which highlights the importance of externalities on the network formation process within communities. Similar to connections across communities, homophily in location, industry and occupation is an important component of the utility of business connections within communities. Coefficients for the edges and externalities terms do not differ greatly depending on whether node covariates are employed for block recovery; however, the importance of homophily when explaining business connections within blocks is higher when the block recovery step does not account for covariates.

\section{Conclusions}\label{section:conclusion}
Networking on the job is an important determinant of mobility and career advancement in many labor markets. In this paper we have studied a network of business relationships using the digital trace of business card exchanges from Eight, a platform for the digitization of business cards containing data from the whole Japan. Our sample contains about 240,000 users of the platform.

Our empirical analysis is guided by a theoretical equilibrium model of network formation where users form relationships based on their preferences for observables, unobservables, and endogenous network features. The stationary equilibrium characterizes the likelihood of observing a network in the data, and we estimate the parameters using approximate maximum likelihood methods. Crucially, the unobserved heterogeneity is discrete, and the equilibrium is a mixture of exponential random graphs \citep{SchweinbergerHandcock2015, Mele2020}. 

We rely on a two-step approach to estimation, first developed in \cite{BabkinEtAl2020}. The first step involves an approximate clustering of the nodes, to estimate the unobserved (discrete) type distribution. The second step estimates the structural payoff parameters using a pseudolikelihood estimatior \citep{BoucherMourifie2017, Snijders2002}.

We propose several algorithmic improvements to the model-based clustering algorithm in \cite{VuEtAl2013}, to include discrete nodal covariates and to speed-up computations through a mix of variational approximations, fast sparse matrix algebra routines and minorization-maximization methods. These improvements allow estimation of the structural model using a massive dataset with about 240,000 users, controlling for (discrete) observable characteristics.

Our analysis shows that this massive business network contains a large number of small (unobserved) business communities. A standard exponential random graph model is unable to capture this feature. This confirms that including unobserved heterogeneity in the network formation model is crucial to understand the business networking patterns in this data.

Our scalable method will allow network researchers to estimate complex models using massive datasets. Previous work on the econometric analysis of large networks has been limited by the complexity of estimation algorithms, and for most studies the definition of a \emph{large network} has been mostly limited to a few thousands of nodes. Our algorithmic improvements makes it possible to analyze networks with hundreds of thousands of nodes, while using relatively few resources.

Furthermore, additional improvements in computational speed and scalability can be obtained, e.g by using GPUs. Larger networks can be handled at a lower cost by means of distributed computing. Crucially, the space complexity of our implementation depends heavily on the size of the matrix of variational parameters, which is a dense matrix and grows with the number of clusters. Since it is reasonable to expect that the number of unobservable blocks grows with the size of the network, the dimensions of this matrix may impose a limitation to the size of networks that can be feasibly analyzed.

We also acknowledge that our implementation of the clustering algorithm makes use of the fact that the network is sparse and the discrete covariates follow the same sparse pattern. In particular, we employ \emph{feature adjacency matrices} to facilitate matrix algebra. Our current implementation requires these matrices to be sufficiently sparse to fit into memory. This complication arises because the algorithm requires the creation of a number of sparse matrices that grows with the square of the number of discrete covariates. Both of these issues impose limitations to the type and number of covariates that can be employed. We believe that it is possible to further improve the speed of the block structure recovery step and at the same time break the dependency on the sparsity of the discrete features, which should make it possible to employ more and better covariates, and to run more iterations of the EM algorithm, thus improving its block recovery capabilities. We expect to extend our algorithm to include these improvements in future versions of this research, and that solving some of this additional complications may contribute to popularize the industrial use of exponential random graph models for the analysis of large social networks and their simulation.

\bibliographystyle{jmr}
\bibliography{biblio}

@article{Mele2017,
    title = {A structural model of dense network formation},
    author = {Angelo Mele},
    journal = {Econometrica},
    volume = {85},
    number = {2},
    year = {2017},
    pages = {825-850}
}

@article{BhamidiEtAl2011,
author = {Shankar Bhamidi and Guy Bresler and Allan Sly},
title = {{Mixing time of exponential random graphs}},
volume = {21},
journal = {The Annals of Applied Probability},
number = {6},
publisher = {Institute of Mathematical Statistics},
pages = {2146-2170},
year = {2011}
}

@Article{AthreyaEtAl2018a,
author = {Athreya, Avanti and Fishkind, Donniell E. and Levin, Keith and Lyzinski, Vince and Park, Youngser and Qin, Yichen and Sussman, Daniel L. and Tang, Mihn and Vogelstein, Joshua T. and Priebe, Carey E.},
title = {Statistical Inference on Random Dot Product Graphs: A Survey},
journal = {Journal of Machine Learning Research},
year = {2018},
volume = {18},
number = {226},
pages = {1-92}
}

@Article{DaudinEtAl2008,
author={Daudin, J.-J.
and Picard, F.
and Robin, S.},
title={A mixture model for random graphs},
journal={Statistics and Computing},
year={2008},
month={Jun},
day={01},
volume={18},
number={2},
pages={173--183}
}

@Unpublished{Mele2020,
author = {Angelo Mele},
title = {A structural model of homophily and clustering in social networks},
month = {May},
year = {2020},
note = {working paper}
}

@Unpublished{MeleEtAl2021,
author = {Mele, Angelo and Hao, Lingxin and Cape, Joshua and Priebe, Carey E.},
title = {Spectral estimation of large stochastic blockmodels with nodal covariates},
month = {February},
year = {2021},
note = {working paper}
}

@ARTICLE{Snijders2002,
  AUTHOR =       {Tom A.B Snijders},
  TITLE =        {Markov Chain Monte Carlo Estimation of
Exponential Random Graph Models},
  JOURNAL =      {Journal of Social Structure},
  YEAR =         {2002},
  volume =       {3},
  number =       {2},
}

@InCollection{DePaula2017,
author = {DePaula, Aureo},
title = {Econometrics of Network Models},
booktitle = {Advances in Economics and Econometrics: Eleventh World Congress},
publisher = {Cambridge University Press},
year = {2017},
editor = {Honore, B.  and Pakes, A. and Piazzesi, M. and Samuelson, L.}
}

@Book{GrahamDePaula2020,
editor = {Graham, Bryan and dePaula, Aureo},
title = {The econometric analysis of network data},
publisher = {Amsterdam: Academic Press, 2020},
year = {2020}
}

@Book{Jackson2008,
editor = {Jackson, Matthew },
title = {Social and economic networks},
publisher = {Princeton},
year = {2008}
}

@InCollection{Graham2020,
author = {Graham, Bryan},
title = {Network data},
publisher = {Amsterdam: North-Holland,},
booktitle = {Handbook of econometrics 7A},
year = {2020},
editor = {Durlauf, S. and Hansen, L. and Heckman,  J. and Matzkin,  R.}
}

@UNPUBLISHED{ChristakisEtAl2010,
  AUTHOR =       {Nicholas Christakis and James Fowler and Guido W. Imbens and Karthik Kalyanaraman},
  TITLE =        {An Empirical Model for Strategic Network Formation},
  NOTE =         {Harvard University},
  year =         {2010},
  month =        {may},
}

@InCollection{Chandrasekhar2016,
author = {Arun G. Chandrasekhar},
editor = {yann bramoulle and andrea galeotti and brian rogers},
booktitle = {Oxford handbook on the economics of networks.},
title = {Econometrics of network formation},
publisher = {Oxford Univerisity Press},
year = {2016}
}

@article{MeleZhu2021,
author = {Mele, Angelo and Zhu, Lingjiong},
title = {Approximate variational estimation for a model of network formation},
journal = {Review of Economics and Statistics},
year = {forthcoming}
}

@article{MondererShapley2006,
author = {Monderer, Dov and Shapley, Lloyd},
title = {Potential Games},
journal = {Games and Economic Behavior},
year = {1996},
volume = {14},
number = {1},
pages = {124-143}
}

@Article{Sheng2020, 
author = {Shuyang Sheng},
title = {A Structural Econometric Analysis of Network Formation Games Through Subnetworks},
journal = {Econometrica},
volume = {88},
number = {5},
pages = {1829-1858},
year = {2020}
}

@Unpublished{BabkinEtAl2020,
author = {Babkin, Sergei and Stewart, Jonathan and Long, Xiaochen and Schweinberger, Michael},
title = {Large-scale estimation of random graph models with local dependence},
note = {working paper, https://arxiv.org/pdf/1703.09301.pdf},
month = {March},
year = {2020}
}

@Article{SchweinbergerStewart2020,
author = {Schweinberger, Michael and Jonathan Stewart},
title = {Concentration and consistency results for canonical and curved exponential-family models of random graphs},
journal = {Annals of Statistics},
year = {2020},
volume = {48},
number = {1},
pages = {374-396.}
}

@Article{Schweinberger2020,
author = {Schweinberger, Michael},
title = {Consistent structure estimation of exponential family random graph models with block structure},
journal = {Bernoulli},
year = {2020},
volume = {26},
pages = {1205–1233}
}

@article{BickelEtAl2013,
author = {Peter Bickel and David Choi and Xiangyu Chang and Hai Zhang},
title = {{Asymptotic normality of maximum likelihood and its variational approximation for stochastic blockmodels}},
volume = {41},
journal = {The Annals of Statistics},
number = {4},
publisher = {Institute of Mathematical Statistics},
pages = {1922 -- 1943},
year = {2013}
}

@Article{SchweinbergerHandcock2015,
	author = {Schweinberger, Michael and Handcock, Mark S },
	title = {Local dependence in random graph models: char- acterization, properties and statistical inference.},
	journal = {Journal of the Royal Statistical Society, Series B (Statistical Methodology)},
	year = {2015},
	number = {77},
	pages = {1-30}
}

@article {BanerjeeEtAl2013,
	author = {Banerjee, Abhijit and Chandrasekhar, Arun G. and Duflo, Esther and Jackson, Matthew O.},
	title = {The Diffusion of Microfinance},
	volume = {341},
	number = {6144},
	year = {2013},
	publisher = {American Association for the Advancement of Science},
	journal = {Science}
}

@Article{BoucherMourifie2017,
author = {Vincent Boucher and Ismael Mourifie},
title = {My Friend Far Far Away: A Random Field Approach to Exponential Random Graph Models},
journal = {Econometrics Journal},
year = {2017},
volume = {20},
number = {3},
pages = {S14-S46}
}

@article{Graham2014,
  AUTHOR =       {Bryan Graham},
  TITLE =        {An empirical model of network formation: with Degree Heterogeneity},
  journal = {Econometrica},
  year =         {2017},
  volume ={85},
  number ={4},
  pages = { 1033-1063}
}

@Article{Leung2015,
  author =       {Michael Leung},
  title =        {Two-Step Estimation of Network-Formation Models with Incomplete Information},
  journal =         {Journal of Ecoometrics},
  year =         {2015},
  volume = {188},
  number = {1},
  pages = {182-195}
}

@Article{Mele2020AEJPol,
title =  {Does School Desegregation Promote Diverse Interactions? An Equilibrium Model of Segregation within Schools},
author = {Angelo Mele},
year = {2020},
journal  = {American Economic Journal: Economic Policy},
volume = {12},
number = {2},
pages = { 228-257}
}

@article{BonhommeLamadonManresa2019,
author = {Bonhomme, Stephane and Lamadon, Thibaut and Manresa, Elena},
title = {A distributional framework for matched employer employee data},
journal = {Econometrica},
volume = {87},
number = {3},
year = {2019},
pages = {699–739}
}

@Article{DePaulaEtAl2014,
  AUTHOR =       {Aureo DePaula and Seth Richards-Shubik and Elie Tamer},
  TITLE =        {Identifying Preferences in Networks with Bounded Degree},
  year =         {2018},
  journal ={Econometrica}, 
  volume = {86}, 
  number = {1},
  pages = {263-288},
  month = {January}
}

@ARTICLE{JacksonWatts2001,
  AUTHOR =       {Matthew Jackson and Alison Watts},
  TITLE =        {The existence of pairwise stable networks},
  JOURNAL =      {Seoul Journal of Economics},
  YEAR =         {2001},
  volume =       {14},
  number =       {3},
  pages =        {299-321}
}

@BOOK{Bishop2006,
  AUTHOR =       {Bishop, Christopher},
  TITLE =        {Pattern recognition and machine learning},
  PUBLISHER =    {Springer},
  YEAR =         {2006},
  address =      {New York}
}

@article{Calvo-ArmengolJackson2004,
author = {Calvo-Armengol, A. and M. O. Jackson},
year = {2004},
title = {The effects of social networks on employment and inequality},
journal = {American Economic Review},
volume = {94}, 
pages = {426–454}
}

@article{BayerRossTopa2008,
author = {Bayer, P. and Ross,  S. and G. Topa},
year = {2008},
title = {Place of work and place of residence: Informal hiring networks and labor market outcomes},
journal = {Journal of Political Economy},
volume = {116}, 
pages = {1150–1196}
}

@article{Galenianos2014,
author = {Manolis Galenianos},
title = {Hiring through Referrals},
journal = {Journal of Economic Theory}, 
year = {2014}, 
volume = {152}, 
pages = {304-323}
}

@article{GaleottiMerlino2014,
author = {Galeotti, Andrea and Merlino, Luca Paolo},
title = {Endogenous job contact networks},
journal = {International Economic Review},
volume = {55},
number = {4},
pages = {1201-1226},
year = {2014}
}

@article{Beaman2012,
author = {Lori Beaman},
title = {Social Networks and the Dynamics of Labor Market Outcomes: Evidence from Refugees Resettled in the U.S.},
journal = {Review of Economic Studies},
volume = {79},
number = {1},
year = {2012},
pages = {128-161}
}

@article{IoannidesLoury2004,
author = {Ioannides, Y. M. and L. D. Loury},
year = {2004},
title = {Job information networks, neighborhood effects, and inequality},
journal = {Journal of Economic Literature},
volume = {42}, 
pages = {1056–1093}
}

@article{CalvoArmengol2004,
author = {Calvo-Armengol, Antoni},
year = {2004},
title = {Job contact networks},
journal = {Journal of Economic Theory},
volume = {115}, 
pages = {191–206}
}

@ARTICLE{WainwrightJordan2008,
  author =       {M.J. Wainwright  and  M.l. Jordan},
  title =        {Graphical  models, exponential families, and variational inference},
  journal =      {Foundations  and  Trends@  in  Machine  Learning},
  year =         {2008},
  volume =       {1},
  number =       {1-2},
  pages =        {1-305}
}

@article{VuEtAl2013,
author = {Duy Q. Vu and David R. Hunter and Michael Schweinberger},
title = {{Model-based clustering of large networks}},
volume = {7},
journal = {The Annals of Applied Statistics},
number = {2},
publisher = {Institute of Mathematical Statistics},
pages = {1010 -- 1039},
year = {2013}
}

@article{Stefanov2004,
  title={Convex quadratic minimization subject to a linear constraint and box constraints},
  author={Stefanov, Stefan M},
  journal={Applied Mathematics Research Express},
  volume={2004},
  number={1},
  pages={17--42},
  year={2004},
  publisher={Hindawi Publishing Corporation}
}

@article{Schweinberger2018,
  title={{HERGM: Hierarchical exponential-family random graph models}},
  author={Schweinberger, Michael and Luna, Pamela},
  journal={Journal of Statistical Software},
  volume={85},
  number={1},
  pages={1--39},
  year={2018}
}

@article{Rosvall_2009,
   title={The map equation},
   volume={178},
   ISSN={1951-6401},
   url={http://dx.doi.org/10.1140/epjst/e2010-01179-1},
   DOI={10.1140/epjst/e2010-01179-1},
   number={1},
   journal={The European Physical Journal Special Topics},
   publisher={Springer Science and Business Media LLC},
   author={Rosvall, M. and Axelsson, D. and Bergstrom, C. T.},
   year={2009},
   month={Nov},
   pages={13–23}
}

@article{Yang2016,
author={Yang, Zhao
and Algesheimer, Ren{\'e}
and Tessone, Claudio J.},
title={A Comparative Analysis of Community Detection Algorithms on Artificial Networks},
journal={Scientific Reports},
year={2016},
month={Aug},
day={01},
volume={6},
number={1},
pages={30750},
issn={2045-2322},
doi={10.1038/srep30750},
url={https://doi.org/10.1038/srep30750}
}

@InProceedings{Pascal2005,
author="Pons, Pascal
and Latapy, Matthieu",
editor="Yolum, pInar
and G{\"u}ng{\"o}r, Tunga
and G{\"u}rgen, Fikret
and {\"O}zturan, Can",
title="Computing Communities in Large Networks Using Random Walks",
booktitle="Computer and Information Sciences - ISCIS 2005",
year="2005",
publisher="Springer Berlin Heidelberg",
address="Berlin, Heidelberg",
pages="284--293",
isbn="978-3-540-32085-2"
}

@article{Acemoglu2012network,
  title={{The Network Origins of Aggregate Fluctuations}},
  author={Acemoglu, Daron and Carvalho, Vasco M and Ozdaglar, Asuman and Tahbaz-Salehi, Alireza},
  journal={Econometrica},
  volume={80},
  number={5},
  pages={1977--2016},
  year={2012},
  publisher={Wiley Online Library}
}

@article{Bernard2019production,
  title={{Production Networks, Geography, and Firm Performance}},
  author={Bernard, Andrew B and Moxnes, Andreas and Saito, Yukiko U},
  journal={Journal of Political Economy},
  volume={127},
  number={2},
  pages={639--688},
  year={2019},
  publisher={The University of Chicago Press Chicago, IL}
}

@Unpublished{Miyauchi2021,
author = {Yuhei Miyauchi},
title = {{Matching and Agglomeration: Theory and Evidence from Japanese Firm-to-Firm Trade}},
month = {April},
year = {2021},
note = {Working Paper}
}

@Unpublished{Konig2018endogenous,
  title={{Endogenous Technology Cycles in Dynamic R\&D Networks}},
  author={K{\"o}nig, Michael D and Rogers, Tim},
  year={2018},
  note={CEPR Discussion Paper No. DP13307}
}

\appendix 




\section{Computational details}\label{appendix:computation}

\subsection{Generalized EM-step: An MM algorithm}
Denote $\bm{g} = [g_{ij}]$ the adjacency matrix of the network.
Recall the surrogate function:
\begin{eqnarray}
M\left(\bm{\xi}; \bm{g},\bm{x},\bm{\alpha},\bm{\beta},\bm{\eta}, \bm{\xi}^{(s)} \right) &:= & \sum_{i<j}^{n} \sum_{k=1}^{K} \sum_{l=1}^{K} \left(\xi_{ik}^{2} \frac{\xi_{jl}^{(s)}}{2\xi_{ik}^{(s)}} + \xi_{jl}^{2} \frac{\xi_{ik}^{(s)}}{2\xi_{jl}^{(s)}}\right) \log \pi_{ij;kl}^{(s)} (g_{ij}, \bm{x},\bm{z}) \notag \\
&+& \sum_{i=1}^{n} \sum_{k=1}^{K} \xi_{ik} \left(\log\eta_{k}^{(s)} - \log\xi_{ik}^{(s)} - \frac{\xi_{ik}}{\xi_{ik}^{(s)}} + 1\right).
\label{eq:minorizer}
\end{eqnarray}
The update rules for $\bm{\xi}$, $\bm{\eta}$, and $\pi_{d_{ij}; x_{ij}}$ follow
\begin{align*}
\bm{\xi}^{(s+1)} &\coloneqq \arg \max_{\bm{\xi}} 
M\left(\bm{\xi}; \bm{g},\bm{x},\bm{\alpha}^{(s)},\bm{\beta}^{(s)},\bm{\eta}^{(s)}, \bm{\xi}^{(s)} \right),
\end{align*}
\begin{align*}
\eta_{k}^{(s+1)} \coloneqq \frac{1}{n} \sum_{i=1}^{n} \xi_{ik}^{(s+1)}, \quad k = 1, \ldots, K,
\end{align*}
and
\begin{eqnarray}
\pi_{ij;kl}^{(s+1)}(d, \chi_{1}, \ldots, \chi_{p},\bm{z}) \coloneqq \frac{\sum_{i=1}^{n}\sum_{j \neq i}\xi_{ik}^{(s+1)} \xi_{jl}^{(s+1)} \bm{1}\lbrace g_{ij}=d, \chi_{1, ij}=\chi_{1}, \ldots, \chi_{p, ij}=\chi_{p} \rbrace }{\sum_{i=1}^{n}\sum_{j \neq i}\xi_{ik}^{(s+1)} \xi_{jl}^{(s+1)} \bm{1}\lbrace \chi_{1, ij}=\chi_{1}, \ldots, \chi_{p, ij}=\chi_{p}\rbrace }
\label{eq:update_pi_appendix}
\end{eqnarray}
for $ k, l = 1, \ldots, K$ and $d, \chi_{1}, \ldots, \chi_{p} \in \{0, 1\} $, respectively.
Maximizing the surrogate function amounts to solving $n$ separate quadratic programming problems of $K$ variables $\bm{\xi_{i}}$ under the constraints  $\xi_{ik} \geq 0$ for all $k$ and $\sum_{k=1}^{n} \xi_{ik} = 1$ \citep{Stefanov2004}.
To do so, we need to compute the coefficients on $\xi_{ik}^{2}$ and $\xi_{ik}$ for all $i$ and $k$.
Note that 
\begin{align*}
   & \sum_{i<j}^{n} \sum_{k=1}^{K} \sum_{l=1}^{K} \left(\xi_{ik}^{2} \frac{\xi_{jl}^{(s)}}{2\xi_{ik}^{(s)}} + \xi_{jl}^{2} \frac{\xi_{ik}^{(s)}}{2\xi_{jl}^{(s)}}\right) \log \pi_{ij;kl}^{(s)} (g_{ij}, \bm{x},\bm{z}) \\
    &= \sum_{i=1}^{n} \sum_{k=1}^{K} \sum_{j \neq i}^{n}  \sum_{l=1}^{K} \frac{\xi_{ik}^{2}}{2\xi_{ik}^{(s)}} \xi_{jl}^{(s)} \log \pi_{ij;kl}^{(s)} (g_{ij}, \bm{x},\bm{z}) \\
    &= \sum_{i=1}^{n} \sum_{k=1}^{K} \frac{\xi_{ik}^{2}}{2\xi_{ik}^{(s)}} \underbrace{\sum_{j \neq i}^{n}  \sum_{l=1}^{K}  \xi_{jl}^{(s)} \log \pi_{ij;kl}^{(s)} (g_{ij}, \bm{x},\bm{z})}_{=: \Omega_{ik}^{(s)}(\bm{g}, \bm{x},\bm{z})} \\
    &= \sum_{i=1}^{n} \sum_{k=1}^{K} \frac{\Omega_{ik}^{(s)}(\bm{g}, \bm{x},\bm{z})}{2\xi_{ik}^{(s)}}\xi_{ik}^{2}.
\end{align*}
Thus the surrogate function can be rearranged as
\begin{align*}
M\left(\bm{\xi}; \bm{g},\bm{x},\bm{\alpha},\bm{\beta},\bm{\eta}, \bm{\xi}^{(s)} \right) = \sum_{i=1}^{n} \sum_{k=1}^{K} \left\{ \frac{1}{\xi_{ik}^{(s)}} \left(\frac{\Omega_{ik}^{(s)}(\bm{g}, \bm{x},\bm{z})}{2\xi_{ik}^{(s)}} - 1 \right)\xi_{ik}^{2} + 
\left(\log\eta_{k}^{(s)} - \log\xi_{ik}^{(s)} + 1\right)  \xi_{ik} \right\}.
\end{align*}

A bottleneck in computation lies in $\Omega_{ik}^{(s)}(\bm{g}, \bm{x},\bm{z}) \coloneqq \sum_{j \neq i}^{n}  \sum_{l=1}^{K}  \xi_{jl}^{(s)} \log \pi_{ij;kl}^{(s)}(g_{ij}, \bm{x},\bm{z})$, since $\pi_{ij;kl}^{(s)}(g_{ij}, \bm{x},\bm{z})$ is dependent on the state of each dyad of the graph.
We could naively compute $\Omega_{ik}^{(s)}(\bm{g}, \bm{x},\bm{z})$ by choosing an appropriate $\pi_{ij;kl}^{(s)}(g_{ij}, \bm{x},\bm{z})$ for each dyad in nested loops, but it would be computationally burdensome.
To overcome this challenge, we will prove that computing $\Omega_{ik}^{(s)}(\bm{g}, \bm{x},\bm{z})$ can be simplified to matrix multiplication and summation by making use of the adjacency matrix of the graph.\footnote{The basic idea had already been implemented in the R package \texttt{hergm} \citep{Schweinberger2018}. One of our contributions is to use an adjacency matrix to speed up the computation.}

To describe how that method works, we start with a simple example where there are no nodal covariates, i.e., we assume $\pi_{ij;kl}^{(s)}(g_{ij}, \bm{x},\bm{z}) = \pi_{ij;kl}^{(s)}(g_{ij}, \bm{z})$ for any $i, j$.
First, we suppose $g_{ij}=0$ for all $i \neq j$ and compute $\Omega_{ik}^{(s)}(\bm{g} = \bm{0}, \bm{z})$. 
Based on the update rule (\ref{eq:update_pi_appendix}), we can drop the notational dependence of $\pi_{ij;kl}^{(s)}(g_{ij}, \bm{z})$ on $i, j$, i.e.,
\begin{align*}
    \pi_{ij;kl}^{(s)}(g_{ij}, \bm{z}) = \pi_{kl}^{(s)}(g_{ij}, \bm{z})
\end{align*}
for all $i, j$.
Observe that
\begin{align*}
     \Omega_{ik}^{(s)}(\bm{g} = \bm{0}, \bm{z}) &= \sum_{j \neq i}^{n}  \sum_{l=1}^{K}  \xi_{jl}^{(s)} \log \pi_{kl}(0, \bm{z}) \\
     &= \sum_{j \neq i}^{n}\left( \xi_{j1}^{(s)} \log \pi_{k1}(0, \bm{z}) + \cdots +  \xi_{jK}^{(s)} \log \pi_{kK}(0, \bm{z}) \right) \\
     &= \left\{ \left( \xi_{11}^{(s)} + \xi_{21}^{(s)} + \cdots + \xi_{n1}^{(s)}\right) - \xi_{i1}^{(s)}  \right\} \log \pi_{k1}(0, \bm{z}) \\
     &+ \left\{ \left( \xi_{12}^{(s)} + \xi_{22}^{(s)} + \cdots + \xi_{n2}^{(s)}\right) - \xi_{i2}^{(s)}  \right\} \log \pi_{k2}(0, \bm{z}) \\
     &+ \cdots \\
     &+ \left\{ \left( \xi_{1K}^{(s)} + \xi_{2K}^{(s)} + \cdots + \xi_{nK}^{(s)}\right) - \xi_{iK}^{(s)}  \right\} \log \pi_{kK}(0, \bm{z}) \\
     &= \sum_{l=1}^{K} \left( \underbrace{\sum_{j=1}^{n} \xi_{jl}^{(s)}}_{=: \tau(l)} - \xi_{il}^{(s)}\right) \log \pi_{kl}(0, \bm{z}) \\
     &= \sum_{l=1}^{K} \left(\tau(l) - \xi_{il}^{(s)}\right)\log \pi_{kl}(0, \bm{z}).
\end{align*}
Define
\begin{align*}
    \underset{(n\times K)}{\bm{A_{0}}} \coloneqq \begin{bmatrix} 
    \tau(1) - \xi_{11}^{(s)} & \tau(2) - \xi_{12}^{(s)} & \dots &  \tau(K) - \xi_{1K}^{(s)} \\
    \tau(1) - \xi_{21}^{(s)} & \tau(2) - \xi_{22}^{(s)}  & \dots & \tau(K) - \xi_{2K}^{(s)} \\
    \vdots & \vdots  & \ddots  & \vdots \\ 
    \tau(1) - \xi_{n1}^{(s)} &  \tau(2) - \xi_{n2}^{(s)} & \dots  & \tau(K) - \xi_{nK}^{(s)}
    \end{bmatrix}
\end{align*}
and
\begin{align*}
    \underset{(K\times K)}{\bm{\Pi_{0}}(\bm{z})} \coloneqq \begin{bmatrix} 
    \log \pi_{11}(0, \bm{z}) & \log \pi_{12}(0, \bm{z}) & \dots & \log \pi_{1K}(0, \bm{z}) \\
    \log \pi_{21}(0, \bm{z}) & \log \pi_{22}(0, \bm{z}) & \dots & \log \pi_{2K}(0, \bm{z}) \\
    \vdots & \vdots  & \ddots  & \vdots \\ 
    \log \pi_{K1}(0, \bm{z}) & \log \pi_{K2}(0, \bm{z}) & \dots  & \log \pi_{KK}(0, \bm{z}).
    \end{bmatrix}
\end{align*}
Then $\Omega_{ik}^{(s)}(\bm{g} = \bm{0}, \bm{z})$ is given by the $(i, k)$ entry of $\bm{A_{0}} \bm{\Pi_{0}}(\bm{z})^{\top}$.

Next, construct the following $K \times K$ matrix:
\begin{align*}
    \underset{(K\times K)}{\bm{\Pi_{1}}(\bm{z})} \coloneqq \begin{bmatrix} 
    \log \frac{\pi_{11}(1, \bm{z})}{\pi_{11}(0, \bm{z})} & \log  \frac{\pi_{12}(1, \bm{z})}{\pi_{12}(0, \bm{z})} & \dots & \log  \frac{\pi_{1K}(1, \bm{z})}{\pi_{1K}(0, \bm{z})} \\
    \log \frac{\pi_{21}(1, \bm{z})}{\pi_{21}(0, \bm{z})} & \log \frac{\pi_{22}(1, \bm{z})}{\pi_{22}(0, \bm{z})} & \dots & \log \frac{\pi_{2K}(1, \bm{z})}{\pi_{2K}(0, \bm{z})} \\
    \vdots & \vdots  & \ddots  & \vdots \\ 
    \log \frac{\pi_{K1}(1, \bm{z})}{\pi_{K1}(0, \bm{z})} & \log \frac{\pi_{K2}(1, \bm{z})}{\pi_{K2}(0, \bm{z})} & \dots  & \log \frac{\pi_{KK}(1, \bm{z})}{\pi_{KK}(0, \bm{z})}.
    \end{bmatrix}
\end{align*}
Remember that $\bm{g} = [g_{ij}]$ is the adjacency matrix of the network, and that $\bm{\xi}^{(s)} = [\xi_{ik}^{(s)}]$ is a $n \times K$ matrix.
We can show that the $(i, k)$ entry of $\bm{g} \bm{\xi}^{(s)} \bm{\Pi_{1}(\bm{z})}^{\top}$ is given by 
\begin{align*}
    \sum_{j \neq i}^{n}  \sum_{l=1}^{K}  g_{ij} \xi_{jl}^{(s)} \log \frac{\pi_{kl}(1, \bm{z})}{\pi_{kl}(0, \bm{z})} = 
     \sum_{j \neq i}^{n}  \sum_{l=1}^{K} (g_{ij} \xi_{jl}^{(s)} \log \pi_{kl}(1, \bm{z}) - g_{ij} \xi_{jl}^{(s)} \log \pi_{kl}(0, \bm{z}))
\end{align*}

Finally, the $(i, k)$ entry of $\bm{A_{0}} \bm{\Pi_{0}}(\bm{z})^{\top} + \bm{g} \bm{\xi}^{(s)} \bm{\Pi_{1}(\bm{z})}^{\top}$ corresponds to
\begin{align*}
   & \sum_{j \neq i}^{n}  \sum_{l=1}^{K}  \xi_{jl}^{(s)} \log \pi_{kl}(0, \bm{z}) + \sum_{j \neq i}^{n}  \sum_{l=1}^{K} (g_{ij} \xi_{jl}^{(s)} \log \pi_{kl}(1, \bm{z}) - g_{ij} \xi_{jl}^{(s)} \log \pi_{kl}(0, \bm{z})) \\ &=
   \sum_{j \neq i}^{n}  \sum_{l=1}^{K} \xi_{jl}^{(s)}(g_{ij} \log \pi_{kl}(1, \bm{z}) + (1 - g_{ij}) \log \pi_{kl}(0, \bm{z})) \\
   &= \sum_{j \neq i}^{n}  \sum_{l=1}^{K}  \xi_{jl}^{(s)} \log \pi_{kl}(g_{ij}, \bm{z}) \\
   &= \Omega_{ik}^{(s)}(\bm{g}, \bm{z}).
\end{align*}
Therefore, computing the coefficient on $\xi_{ik}^{2}$ of the surrogate function can reduce to matrix multiplication and summation.
Intuitively, assuming that there are no links in the network, we first compute the ``baseline" quadratic coefficients.
Then we update the baseline quadratic coefficients for actual links using the adjacency matrix.\footnote{We can also embed sparse matrix algebra routines for the adjacency matrix, which reduces both the computing time and the burden on the memory usage.}
As Table \ref{tab:benchmark_quadratic_coef} shows, the computing time of $\Omega_{ik}^{(s)}(\bm{g}, \bm{x}, \bm{z})$ using this formula is more than 14,000 times faster than that of a naive way using nested loops.
This helps reduce the computing time in the overall EM steps, since the quadratic coefficients need to be computed in every EM iteration.

\begin{table}[h!]
\caption{Comparison of the computing time of quadratic coefficients between nested loops and the matrix formula. For this benchmark we used a simulated network with $n = 1,000$ and $K = 50$. Both are calculated using the \texttt{Rcpp} package in \texttt{R}.}
\label{tab:benchmark_quadratic_coef}
\begin{center}
\begin{tabular}{lrr}
\toprule
Method & Time (seconds) & Relative\\
\midrule
Nested loops & 318.590 & 14481.36\\
Using the formula & 0.022 & 1.00\\
\bottomrule
\end{tabular}
\end{center}
\end{table}

Once the quadratic coefficients are computed, the other steps are relatively easy to proceed.
Computing the coefficients on the linear term in the surrogate function is straightforward, and we can solve the quadratic programming problems with the computed coefficients on the quadratic  and linear terms.
This maximization converges quite fast even for large $n$ and $K$ and is parallelizable.\footnote{Theoretically, this maximization can be conducted in parallel, but we do not employ parallel computing here due to issues related to the OpenMP library. However, the computation converges quite fast even if we do not parallelize this step.}
Based on the $\xi^{(s+1)}$ obtained from the quadratic programming problems, we can update the $K \times K$ matrix $\bm{\pi}(d=1)$ by
\begin{align*}
    \bm{\pi}^{(s+1)}(d=1) = \left\{\left(\bm{\xi}^{(s+1)}\right)^{\top} \bm{g} \left(\bm{\xi}^{(s+1)}\right) \right\} \oslash \left\{\left(\bm{\xi}^{(s+1)} \right)^{\top} \left(\bm{\xi}^{(s+1)}\right) \right\},
\end{align*}
where $\bm{A} \oslash \bm{B}$ stands for the Hadamard (entry-wise) division of the conformable matrices $\bm{A}$ and $\bm{B}$.

So far we consider a special case where there are no nodal covariates in the MM algorithm, but this can be extended to include discrete ones.
We first calculate
\begin{eqnarray}
\pi_{kl}^{(s+1)}(d=1, \chi_{1}=0, \ldots, \chi_{p}=0,\bm{z}) \coloneqq \frac{\sum_{i=1}^{n}\sum_{j \neq i}\xi_{ik}^{(s+1)} \xi_{jl}^{(s+1)} \bm{1}\lbrace g_{ij}=1, \chi_{1, ij}=0, \ldots, \chi_{p, ij}=0 \rbrace }{\sum_{i=1}^{n}\sum_{j \neq i}\xi_{ik}^{(s+1)} \xi_{jl}^{(s+1)} \bm{1}\lbrace \chi_{1, ij}=0, \ldots, \chi_{p, ij}=0\rbrace }
\label{eq:pi_covariate}
\end{eqnarray}
for all $k, l = 1, \ldots, K$. 
This is the probability that nodes in block $k$ and $l$ are connected on average, given that each dyad does not share any characteristics.
Once $\pi_{kl}^{(s+1)}(d=1, \chi_{1}=0, \ldots, \chi_{p}=0,\bm{z})$ is ready, $\pi_{kl}^{(s+1)}(d=0, \chi_{1}=0, \ldots, \chi_{p}=0,\bm{z})$ can be easily computed by $1 - \pi_{kl}^{(s+1)}(d=1, \chi_{1}=0, \ldots, \chi_{p}=0,\bm{z})$, and $\pi_{kl}^{(s+1)}(d=0, \chi_{1}=0, \ldots, \chi_{p}=0,\bm{z})$ will be used to compute the ``baseline" quadratic coefficients as in the case without nodal covariates.

To compute $\pi_{kl}^{(s+1)}(d=1, \chi_{1}=0, \ldots, \chi_{p}=0,\bm{z})$, we use \textit{feature adjacency matrices}, which represent whether node $i$ and $j$ such that $i \neq j$ share the same characteristics (e.g., whether they belong to the same industry).
Define the $n \times n$ feature adjacency matrix for covariate $s$ as
\begin{align*}
    X_{s, ij} = 
    \begin{cases}
    1 & \text{if } s_{i} = s_{j} \text{ and } i \neq j \\
    0 & \text{otherwise}.
    \end{cases}
\end{align*}
Then the $K \times K$ matrix $\bm{\pi}^{(s+1)}(d=1, \chi_{1}=0, \ldots, \chi_{p}=0,\bm{z})$ can be given by
\begin{align*}
    \bm{\pi}^{(s+1)}(d=1, \chi_{1}=0, \ldots, \chi_{p}=0,\bm{z}) = \left\{\left(\bm{\xi}^{(s+1)}\right)^{\top} \bm{g} \circ (\bm{J} - \bm{X_{1}}) \circ \cdots \circ (\bm{J} - \bm{X_{p}})\ \left(\bm{\xi}^{(s+1)}\right) \right\} \oslash \\
    \left\{\left(\bm{\xi}^{(s+1)} \right)^{\top} (\bm{J} - \bm{X_{1}}) \circ \cdots \circ (\bm{J} - \bm{X_{p}}) \left(\bm{\xi}^{(s+1)} \right)\right\},
\end{align*}
where $\bm{A} \circ \bm{B}$ denotes the Hadamard (i.e., entry-wise) product of the conformable matrices $\bm{A}$ and $\bm{B}$, $\bm{J$} is a $n \times n$ matrix whose off-diagonal entries are all one and whose diagonals are all zero.
Note that the sparseness of $\bm{g} \circ (\bm{J} - \bm{X_{1}}) \circ \cdots \circ (\bm{J} - \bm{X_{p}})$ does not exceed that of $\bm{g}$. 
As long as $\bm{g}$ is sparse, $\bm{g} \circ (\bm{J} - \bm{X_{1}}) \circ \cdots \circ (\bm{J} - \bm{X_{p}})$ is also sparse.
This is why we first compute $\bm{\pi}^{(s+1)}(d=1, \chi_{1}=0, \ldots, \chi_{p}=0,\bm{z})$ instead of $ \bm{\pi}^{(s+1)}(d=0, \chi_{1}=0, \ldots, \chi_{p}=0,\bm{z})$, whose numerator contains only dense matrices.
Also note that the matrix $(\bm{J} - \bm{X})$ is very dense when $\bm{X}$ is sparse, and computing naively $(\bm{J} - \bm{X})$ would cause extremely high memory usage for large networks.
This computation is avoidable by using the distributive property
\begin{align*}
   \bm{g} \circ (\bm{J} - \bm{X}) = \bm{g} - \bm{g} \circ \bm{X}
\end{align*}
and applying this property iteratively.

That is not the case with the denominator of $\bm{\pi}^{(s+1)}(d=1, \chi_{1}=0, \ldots, \chi_{p}=0,\bm{z})$, since it only contains dense matrices $(\bm{J} - \bm{X_{1}}), \ldots$, $(\bm{J} - \bm{X_{p}})$.
We want to compute $\left(\bm{\xi}^{(s+1)} \right)^{\top} (\bm{J} - \bm{X_{1}}) \circ \cdots \circ (\bm{J} - \bm{X_{p}}) \left(\bm{\xi}^{(s+1)} \right)$ without breaking matrix sparseness.
To explain how to achieve the objective, let us focus on a case where there are only three feature adjacency matrices $\bm{X_{1}}$, $\bm{X_{2}}$, and $\bm{X_{3}}$.
Thus we consider to calculate $\left(\bm{\xi}^{(s+1)} \right)^{\top} (\bm{J} - \bm{X_{1}}) \circ (\bm{J} - \bm{X_{2}}) \circ (\bm{J} - \bm{X_{3}}) \left(\bm{\xi}^{(s+1)} \right)$.

Note that the following decomposition holds:
\begin{align*}
   &\left(\bm{\xi}^{(s+1)} \right)^{\top} (\bm{J} - \bm{X_{1}}) \circ (\bm{J} - \bm{X_{2}}) \circ (\bm{J} - \bm{X_{3}}) \left(\bm{\xi}^{(s+1)} \right) \\
   &= \left(\bm{\xi}^{(s+1)} \right)^{\top} \left\{\bm{J} - (\bm{X_{1}} + \bm{X_{2}} + \bm{X_{3}}) + (\bm{X_{1}} \circ \bm{X_{2}} + \bm{X_{2}} \circ \bm{X_{3}} + \bm{X_{3}} \circ \bm{X_{1}}) - (\bm{X_{1}} \circ \bm{X_{2}} \circ \bm{X_{3}}) \right\} \left(\bm{\xi}^{(s+1)} \right) \\
   &= \underbrace{\left(\bm{\xi}^{(s+1)} \right)^{\top} \bm{J} \left(\bm{\xi}^{(s+1)} \right)}_{=:\bm{P_{1}}} \\
   &+ \underbrace{\left(\bm{\xi}^{(s+1)} \right)^{\top} \left\{ - (\bm{X_{1}} + \bm{X_{2}} + \bm{X_{3}}) + (\bm{X_{1}} \circ \bm{X_{2}} + \bm{X_{2}} \circ \bm{X_{3}} + \bm{X_{3}} \circ \bm{X_{1}}) - (\bm{X_{1}} \circ \bm{X_{2}} \circ \bm{X_{3}}) \right\} \left(\bm{\xi}^{(s+1)} \right)}_{=:\bm{P_{2}}} \\
   &= \bm{P_{1}} + \bm{P_{2}}.
\end{align*}
$(\bm{X_{1}} + \bm{X_{2}} + \bm{X_{3}})$,  $(\bm{X_{1}} \circ \bm{X_{2}} + \bm{X_{2}} \circ \bm{X_{3}} + \bm{X_{3}} \circ \bm{X_{1}})$, and $(\bm{X_{1}} \circ \bm{X_{2}} \circ \bm{X_{3}})$ are all sparse as long as $\bm{X_{1}}$, $\bm{X_{2}}$, and $\bm{X_{3}}$ are sparse. 
Then computing $\bm{P_{2}}$ is not so computationally costly.
Moreover, we can compute $\bm{P_{1}}$ utilizing the fact that the $(i, j)$ entry of the $K \times n$ matrix $(\bm{\xi}^{(s+1)})^{\top} \bm{J}$ equals $\sum_{m \neq j} \xi_{mi}$.
In this way, we can avoid computing dense matrix to compute the denominator of (\ref{eq:pi_covariate}).
So far we have focused on the three-covariate case, and this can be easliy extended to the $p$-covariate one.

Once computing $\bm{\pi}$ is completed, we can calculate the quadratic coefficients as we do without nodal covariates. Redefine
\begin{align*}
    \underset{(n\times K)}{\bm{A_{0}}} \coloneqq \begin{bmatrix} 
    \tau(1) - \xi_{11}^{(s)} & \tau(2) - \xi_{12}^{(s)} & \dots &  \tau(K) - \xi_{1K}^{(s)} \\
    \tau(1) - \xi_{21}^{(s)} & \tau(2) - \xi_{22}^{(s)}  & \dots & \tau(K) - \xi_{2K}^{(s)} \\
    \vdots & \vdots  & \ddots  & \vdots \\ 
    \tau(1) - \xi_{n1}^{(s)} &  \tau(2) - \xi_{n2}^{(s)} & \dots  & \tau(K) - \xi_{nK}^{(s)}
    \end{bmatrix},
\end{align*}
\begin{align*}
    \underset{(K\times K)}{\bm{\Pi_{0}}(\bm{z})} \coloneqq \begin{bmatrix} 
    \log \pi_{d=0, chi_{1}=0, \ldots, \chi_{p}=0;11} & \log \pi_{d=0, \chi_{1}=0, \ldots, \chi_{p}=0;12} & \dots & \log \pi_{d=0, \chi_{1}=0, \ldots, \chi_{p}=0;1K} \\
    \log \pi_{d=0, \chi_{1}=0, \ldots, \chi_{p}=0;21} & \log \pi_{d=0, \chi_{1}=0, \ldots, \chi_{p}=0;22} & \dots & \log \pi_{d=0, \chi_{1}=0, \ldots, \chi_{p}=0;2K} \\
    \vdots & \vdots  & \ddots  & \vdots \\ 
    \log \pi_{d=0, \chi_{1}=0, \ldots, \chi_{p}=0;K1} & \log \pi_{d=0, \chi_{1}=0, \ldots, \chi_{p}=0;K2} & \dots  & \log \pi_{d=0, \chi_{1}=0, \ldots, \chi_{p}=0;KK}
    \end{bmatrix},
\end{align*}
and
\begin{align*}
    \underset{(K\times K)}{\bm{\Pi}(d, \chi_{1}, \ldots, \chi_{p}, \bm{z})} \coloneqq \begin{bmatrix} 
    \log \frac{\pi_{d, \chi_{1}, \ldots, \chi_{p};11}}{\pi_{d=0, \chi_{1}=0, \ldots, \chi_{p}=0;11}} & \log  \frac{\pi_{d, \chi_{1}, \ldots, \chi_{p};12}}{
    \log \pi_{d=0, \chi_{1}=0, \ldots, \chi_{p}=0;12}} & \dots & \log  \frac{\pi_{d, \chi_{1}, \ldots, \chi_{p};1K}}{\pi_{d=0, \chi_{1}=0, \ldots, \chi_{p}=0;1K}} \\
    \log \frac{\pi_{d, \chi_{1}, \ldots, \chi_{p};21}}{\pi_{d=0, \chi_{1}=0, \ldots, \chi_{p}=0;21}} & \log \frac{\pi_{d, \chi_{1}, \ldots, \chi_{p};22}}{\pi_{d=0, \chi_{1}=0, \ldots, \chi_{p}=0;22}} & \dots & \log \frac{\pi_{d, \chi_{1}, \ldots, \chi_{p};2K}}{\pi_{d=0, \chi_{1}=0, \ldots, \chi_{p}=0;2K}} \\
    \vdots & \vdots  & \ddots  & \vdots \\ 
    \log \frac{\pi_{d, \chi_{1}, \ldots, \chi_{p};K1}}{\pi_{d=0, \chi_{1}=0, \ldots, \chi_{p}=0;K1}} & \log \frac{\pi_{d, \chi_{1}, \ldots, \chi_{p};K2}}{\pi_{d=0, \chi_{1}=0, \ldots, \chi_{p}=0;K2}} & \dots  & \log \frac{\pi_{d, \chi_{1}, \ldots, \chi_{p};KK}}{\pi_{d=0, \chi_{1}=0, \ldots, \chi_{p}=0;KK}}
    \end{bmatrix}.
\end{align*}
Also define functions $\bm{\Gamma} \colon \{0, 1\} \to \mathbb{R}^{n \times n}$ and $\bm{\Lambda}_{s} \colon \{0, 1\} \to \mathbb{R}^{n \times n}$ $(s = 1, \ldots, p)$ such that
\begin{align*}
    \bm{\Gamma}(d) \coloneqq 
    \begin{cases}
    \bm{G} & (d = 1) \\
    \bm{J} - \bm{G} & (d = 0)
    \end{cases}
\end{align*}
and
\begin{align*}
    \bm{\Lambda}_{s}(\chi_{s}) \coloneqq
    \begin{cases}
    \bm{X}_{s} & (\chi_{s} = 1) \\
    \bm{J} - \bm{X}_{s} &  (\chi_{s} = 0)
    \end{cases}.
\end{align*}
Then $\Omega_{ik}^{(s)}(\bm{g}, \bm{x},\bm{z}) \coloneqq \sum_{j \neq i}^{n}  \sum_{l=1}^{K}  \xi_{jl}^{(s)} \log \pi_{kl}^{(s)}(g_{ij}, \bm{x},\bm{z})$, the most time-consuming part to compute, is given by the $(i, k)$ entry of the following matrix:
\begin{align*}
    \bm{A_{0}}\bm{\Pi_{0}}(\bm{z})^{\top} + \sum_{d = \chi_{1}=\cdots=\chi_{p} \neq 0} \bm{\Gamma}(d) \circ \bm{\Lambda_{1}}(\chi_{1}) \circ \cdots \circ \bm{\Lambda_{p}}(\chi_{p}) \bm{\xi}^{(s)} \bm{\Pi}(d, \chi_{1}, \ldots, \chi_{p})^{\top}.
\end{align*}
The rest of the computation proceeds in the same way as we do without nodal covariates.

\end{document}